\documentclass[sigconf]{acmart}

\AtBeginDocument{%
  }

\setcopyright{acmlicensed}
\copyrightyear{2018}
\acmYear{2018}
\acmDOI{XXXXXXX.XXXXXXX}
\acmConference[Conference acronym 'XX]{Make sure to enter the correct
  conference title from your rights confirmation email}{June 03--05,
  2018}{Woodstock, NY}
\acmISBN{978-1-4503-XXXX-X/2018/06}

\usepackage{graphicx}
\usepackage{booktabs}
\usepackage{multicol,multirow}
\usepackage{amsmath}
\usepackage{amsthm}
\usepackage{threeparttable}
\usepackage{enumitem}
\usepackage{algorithm}
\usepackage[noend]{algpseudocode}

\usepackage{xcolor} 
\definecolor{lightgray}{rgb}{0.95, 0.95, 0.95}
\usepackage{listings}            
\begin{document}

\title{Give Users the Wheel: Towards Promptable Recommendation Paradigm}

\author{Fuyuan Lyu}
\authornote{Both authors contributed equally to this research.}
\email{fuyuan.lyu@mail.mcgill.ca}
\affiliation{%
  \institution{McGill \& Mila - Quebec AI Institute}
  \city{Montreal}
  \country{Canada}
}
\orcid{0000-0001-9345-1828}

\author{Chenglin Luo}
\authornotemark[1]
\email{chenglinluoo@gmail.com}
\affiliation{%
  \institution{Shenzhen Technological University}
  \city{Shenzhen}
  \country{China}
}
\orcid{0009-0008-3958-3123}

\author{Qiyuan Zhang}
\email{qzhang732-c@my.cityu.edu.hk}
\affiliation{%
  \institution{City Univerisity of Hong Kong}
  \city{Hong Kong}
  \country{China}
}
\orcid{0009-0006-1397-085X}

\author{Yupeng Hou}
\email{yupenghou97@gmail.com}
\affiliation{%
  \institution{University of California San Diego}
  \city{San Diego}
  \country{US}
}
\orcid{0000-0002-0747-8010}

\author{Haolun Wu}
\email{haolun.wu@mail.mcgill.ca}
\affiliation{%
  \institution{McGill \& Mila - Quebec AI Institute}
  \city{Montreal}
  \country{Canada}
}
\orcid{0000-0002-8206-4969}

\author{Xing Tang}
\email{xing.tang@hotmail.com}
\affiliation{%
  \institution{Shenzhen Technological University}
  \city{Shenzhen}
  \country{China}
}
\orcid{0000-0003-4360-0754}

\author{Xue Liu}
\email{xue.liu@cs.mcgill.ca}
\affiliation{%
  \institution{McGill \& MBZUAI}
  \city{Abu Dhabi}
  \country{UAE}
}
\orcid{0000-0001-5252-3442}

\author{Jin L.C. Guo}
\email{jguo@cs.mcgill.ca}
\affiliation{%
  \institution{McGill University}
  \city{Montreal}
  \country{Canada}
}
\orcid{0000-0003-1782-1545}

\author{Xiuqiang He}
\email{he.xiuqiang@gmail.com}
\affiliation{%
  \institution{Shenzhen Technological University}
  \city{Shenzhen}
  \country{China}
}
\orcid{0000-0002-4115-8205}

\renewcommand{\shortauthors}{Fuyuan Lyu et al.}

\newcommand{\fuyuan}[1]{\textcolor{blue}{[Fuyuan: #1]}}

\begin{abstract}
Conventional sequential recommendation models have achieved remarkable success in mining implicit behavioral patterns. 
However, these architectures remain structurally blind to explicit user intent: they struggle to adapt when a user's immediate goal (e.g., expressed via a natural language prompt) deviates from their historical habits.
While Large Language Models (LLMs) offer the semantic reasoning to interpret such intent, existing integration paradigms force a dilemma:
LLM-as-a-recommender paradigm sacrifices the efficiency and collaborative precision of ID-based retrieval, while Reranking methods are inherently bottlenecked by the recall capabilities of the underlying model.
In this paper, we propose Decoupled Promptable Sequential Recommendation (DPR), a model-agnostic framework that empowers conventional sequential backbones to natively support Promptable Recommendation, the ability to dynamically steer the retrieval process using natural language without abandoning collaborative signals. DPR modulates the latent user representation directly within the retrieval space.
To achieve this, we introduce a Fusion module to align the collaborative and semantic signals, a Mixture-of-Experts (MoE) architecture that disentangles the conflicting gradients from positive and negative steering, and a three-stage training strategy that progressively aligns the semantic space of prompts with the collaborative space. 
Extensive experiments on real-world datasets demonstrate that DPR significantly outperforms state-of-the-art baselines in prompt-guided tasks while maintaining competitive performance in standard sequential recommendation scenarios.
We will open-source all artifacts after acceptance.
\end{abstract}



\maketitle

\section{Introduction}

Recommendation systems (RS) serve as the fundamental engine of modern online platforms, designed to anticipate user needs via mining historical behavior. Conventional architectures, such as SASRec~\citep{SASRec} and GRU4Rec~\citep{GRU4Rec}, primarily rely on users' behaviour sequences to capture complex temporal patterns and collaborative signals. However, a critical disconnect exists in this paradigm: {while these models achieve superior precision in exploiting latent historical patterns, they remain fundamentally oblivious to the intrinsic dynamism of real-time user intent}. Consider a straightforward example: an action movie fan explicitly requests some children's movies for an evening with kids. A conventional model, bound by the inertia of historical data, would rigidly continue pushing thrillers, ignoring the user’s immediate request. Crucially, the majority of recommendation systems lack the capability to accommodate users' direct instructions, leaving the most accurate signal of intent completely underutilized.

\begin{figure}[!htbp]
    \centering
    \vspace{-10pt}
    \includegraphics[width=0.98\linewidth]{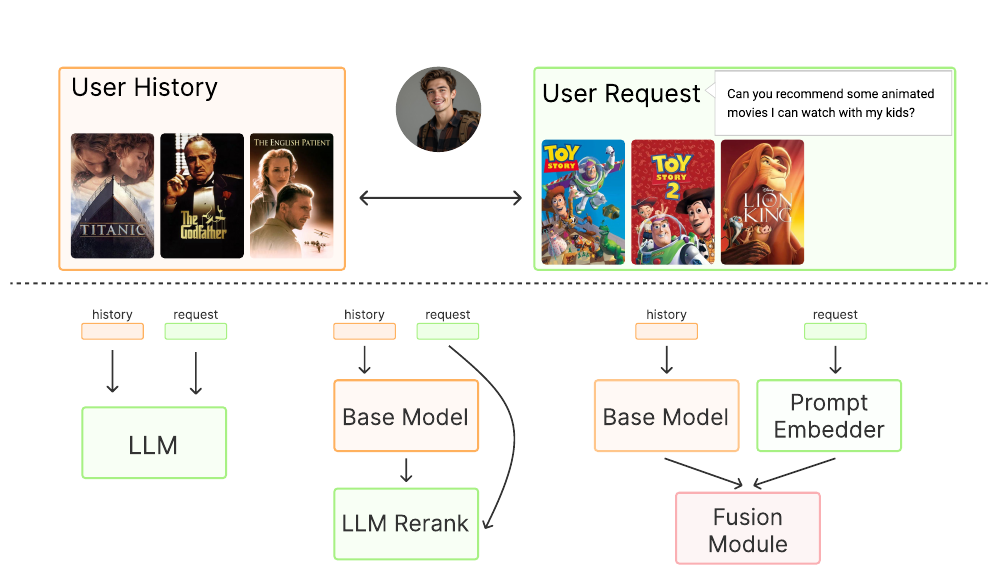}
    \vspace{-10pt}
    \caption{Illustration Figure on How to follow the user's instruction while considering historical interests}
    \vspace{-5pt}
    \Description{Illustration Figure on How to follow the user's instruction while considering historical interests}
    \label{fig:illustration}
\end{figure}

Large Language Models (LLMs) have emerged as a promising tool for interpreting such nuanced preferences due to their strong semantic reasoning capabilities~\cite{LLM-CF}. However, existing approaches to integrating LLMs into recommendations typically follow two main paradigms, each with its own limitations:
{First}, the LLM-as-recommender paradigm directly replaces a conventional recommender with a LLM~\cite{Laser,CTRL-Rec}. While semantically powerful, this approach discards the very essence of collaborative filtering, the large-scale, rich, fine-grained item-item and user-user signals. Although certain works aim to mitigate such drawbacks by finetuning collaborative signals into LLMs, they inevitably suffer from inference latency, rendering them impractical for high-throughput systems.
{Second}, the Reranking paradigm tends to use LLMs as a plug-in re-ranker rather than making the final recommendation directly. However, these works~\cite{UR4Rec,QueRec} are fundamentally shackled by the initial ranking stage. If the conventional model fails to recall relevant items due to a shift in user intent (e.g., a sudden demand for "comedy"), the LLM re-ranker has no valid candidates to optimize. Ultimately, current research forces a dilemma: one must choose between the collaborative precision of ID-based models and the instruction-following capability of LLMs. Their illustrations are shown in Figure~\ref{fig:illustration}.

In this paper, we challenge this enforced trade-off by asking a pivotal question: \textbf{Is it possible to "prompt" a conventional recommendation model directly through natural language?} 
We argue that the ideal recommendation system should not replace the efficient conventional backbone, nor should it merely patch it with a re-ranker. Instead, it should intrinsically modulate the retrieval mechanism itself.
We formalize this vision as Promptable Recommendation:
\begin{definition}[\textit{Promptable Recommendation}]
A Promptable Recommendation system is a recommender that natively integrates natural language prompts into the collaborative retrieval process. It retains the efficiency of vector-based retrieval and uses explicit user intent to \textit{dynamically steer} the entire search space, ensuring that the retrieved candidates align with both the users' real-time instructions and their historical habits.
\end{definition}

To effectively initiate a prompt recommender, we identify three technical challenges to be addressed. 
The first issue is the semantic-collaborative alignment problem of mapping discrete, open-ended natural language prompts into the continuous, high-dimensional latent space of existing recommendation models without disrupting the manifold of collaborative signals.
The second issue lies in the complexity of user requests. Users' requests may range from positive steering (e.g., "show me comedy") to negative constraints (e.g., "no horror"). A promptable model must be capable of simultaneously amplifying desired features and unlearning restricted attributes within the representation space. 
Finally, the framework must exhibit robustness across modalities. The system is required to function seamlessly whether a prompt is present or absent. Explicit instructions, namely the user prompt, should act as an additive signal that enhances specific queries without disrupting the foundational collaborative filtering performance during passive usage.

To address the aforementioned challenges, we propose the Decoupled Promptable Sequential Recommendation (DPR) framework. DPR introduces a model-agnostic architecture that seamlessly equips existing sequential backbones (e.g., SASRec, BERT4Rec) with promptability. 
To effectively align the semantic and collaborative signals, we propose a fusion module that uses multi-head attention to seamlessly integrate them.
To master the complexity of user intent, we design a Mixture-of-Experts (MoE) Tower architecture that disentangles the distinct optimization dynamics of positive steering and negative unlearning, enabling the system to effectively amplify desired features and suppress restricted attributes by routing requests to specialized experts. 
Furthermore, to ensure robustness across diverse modalities, we implement a Three-Stage Training Strategy. This progressive curriculum aligns latent user representations with textual semantics, ensuring the model functions reliably whether prompts are present or absent. 
Extensive experiments across real-world datasets and various experimental setups demonstrate that DPR significantly outperforms existing methods, including both LLM-based and conventional model-based methods, on prompting tasks while maintaining competitive performance in standard sequential recommendation scenarios. 
Our main contributions are summarized as follows: 
\begin{itemize}[leftmargin=*]
    \item We define the Promptable Recommendation paradigm, which combines the efficiency of collaborative filtering with the controllability of natural language, enabling users to actively steer recommendations via positive steering and negative constraints. 
    \item We propose the DPR framework, a model-agnostic solution that ensures architectural universality through a decoupled design. It features a fusion module to align semantic and collaborative signals, a Mixture-of-Experts (MoE) Tower that specializes in handling diverse user requests, and a three-stage training strategy that handles diverse modalities.
    \item Extensive experiments on real-world datasets demonstrate the effectiveness of DPR under various setups. 
\end{itemize}








\section{Related Work}

\subsection{Sequential Recommendation: from conventional to LLM-powered}
Sequential recommendation (SR) has emerged as a cornerstone of modern information retrieval. Early works~\cite{FPMC,Fossil} adopt statistical models with an MF term and an item-item transition term to capture long and short-term preferences. With the improvement of deep learning, various architectures have been adopted in sequential recommendation. For example, GRU4Rec~\cite{GRU4Rec} adopts the Gated Recurrent Unit to model click sequences for session-based recommendation. Attention-based architecture, such as self-attention module~\cite{SASRec,NARM}, convolutional module~\cite{Caser}, BERT~\cite{BERT4Rec}, and diffusion model~\cite{Diffurec}, have been introduced to capture user interest. After the emergence of LLMs, various models have been leveraging their inherent knowledge~\cite{li2024large,xu2025tapping,bao2023large}. One line of work tend to utilize LLM in an offline manner, such as encoding textual features~\cite{LLM-ESR,LLM4SeqRec} or enhance training~\cite{LLM-CF,LLMSeR}. The other line of work aims to transform a pre-trained LLM into a sequential recommender~\cite{P5,CR-ALign,SeLLa-Rec,RLMRec}. Our proposed paradigm lies in the middle, aiming to leverage the LLM's ability to extract the user's explicit request while maintaining compatibility with the current sequential recommendation pipeline.

\subsection{Natural Language Control over Recommendation}
With the empowerment of LLM, natural language has been utilized as a controller over existing models due to its nature. LLM itself poses such power when presented with user's historical information.
For example, prior work on instruction-following and conversational recommendation models~\cite{vats2024exploring,SeLLa-Rec,P5,zhang2025instructrec,he2023crs} builds on LLMs to introduce steerable capabilities for recommendation. However, the collaborative signals from the user-item interaction history are difficult for LLMs to learn~\cite{he2025reindex}, and these approaches often suffer from efficiency issues~\cite{lin2025efficiency}.
CTRL-Rec~\cite{CTRL-Rec}, on the other hand, utilizes two separate LLMs to compute the representation of both user's request and items' features. However, it limits the interaction format to matrix factorization, neglecting more complex sequential recommendation solutions. Our work aligns with the intent to control recommendation results via natural language~\cite{google-dpr}.


\section{Methodology}
\label{sec:methodology}

In this section, we present the \textbf{Decoupled Promptable Sequential Recommendation (DPR)} framework. In Section~\ref{sec:method_formulation}, we formulate both the sequential and promptable recommendation. Section~\ref{sec:method_arch} describes the decoupled model architecture, consisting of a base sequential encoder and a dual-path prompt fusion module. Finally, Section~\ref{sec:method_train} details our unified training objective and the multi-stage fine-tuning strategy.


\subsection{Problem Formulation} \label{sec:method_formulation}
\paragraph{Sequential Recommendation} Let $\mathcal{U}$ and $\mathcal{V}$ denote the set of users and items, respectively. For a specific user $u \in \mathcal{U}$, their historical interaction sequence is represented as $S_u = [v_1, v_2, \dots, v_t]$, where $v_i \in \mathcal{V}$. The target of sequential recommendation is to predict the next item $v_{t+1}$, formulated as: $\hat{v}_{t+1} = \max_{v}P(v | S_u)$. Losses such as cross-entropy are commonly adopted to measure the distance between $v_{t+1}$ and $\hat{v}_{t+1}$. Hence, the training objective can be derived as 
\begin{equation} \label{eq:seq}
    \min \mathcal{L}_{seq}, \mathcal{L}_{seq} = \sum_{u,t} l_{CE}(\hat{v}_{t+1}, v_{t+1}).
\end{equation}

\paragraph{Promptable Recommendation} The promptable recommendation task introduces a natural language instruction, denoted as the prompt $p$. Associated with $p$ is a semantic indicator $c \in \{+, -\}$, which can be easily inferred from the semantic intent of $p$, as shown in Appendix \ref{appendix:intent_classification}. Here, we categorize the intent $p$ into two classes, where $c = +$ signifies a positive desire (e.g., \textit{``I want...''}) and $c = -$ signifies a negative constraint (e.g., \textit{``Do not show...''}). Our objective is to ensure the output $\hat{v}_{t+1} = \max_{v}P(v | S_u, p, c)$ aligns with both the historical interests $S_u$ and the immediate instruction $p$.

\subsection{Model Architecture} \label{sec:method_arch}
The DPR framework comprises two decoupled components: a \textbf{Sequential Encoder} for extracting historical interests, a \textbf{Prompt Embedder} to encode instructions, and a \textbf{Signal Fusion Module} for aligning semantic and collaborative signals. 
\begin{figure}[!htbp]
    \centering
    \vspace{-5pt}
    \includegraphics[width=0.9\linewidth]{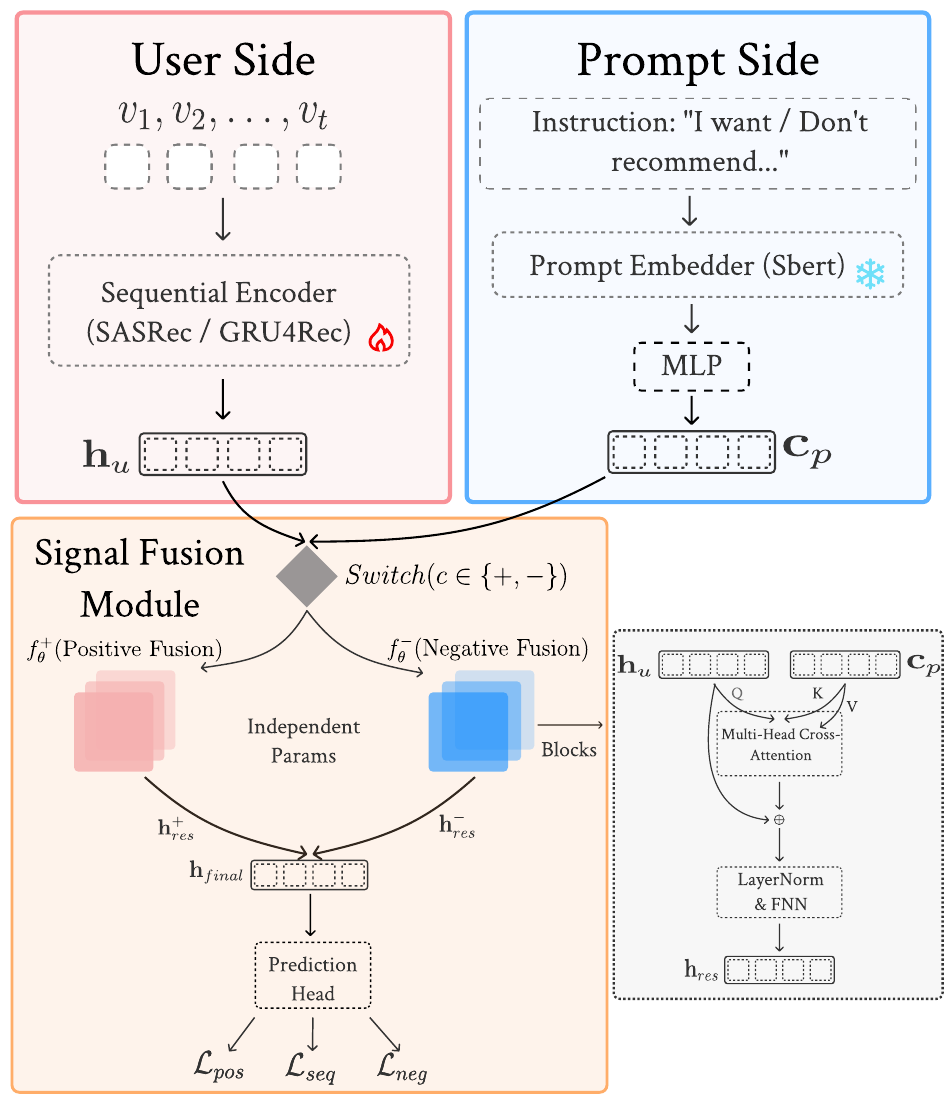}
    \vspace{-10pt}
    \caption{Overall Framework}
    \Description{Overall Framework}
    \vspace{-10pt}
    \label{fig:framework}
\end{figure}

\subsubsection{Sequential Encoder}
To capture long-term user preferences, we employ a sequential encoder $f_\theta$. It is important to note that our framework is \textit{model-agnostic}, hence $f_\theta$ can be instantiated using various architectures, such as attention-based models (e.g., SASRec~\cite{SASRec}, BERT4Rec~\cite{BERT4Rec}) or RNN-based models (e.g., GRU4Rec~\cite{GRU4Rec}).
Given the input sequence $S_u$, the encoder outputs the user's intrinsic interest representation $\mathbf{h}_u \in \mathbb{R}^d$:
\begin{equation}
    \mathbf{h}_u = f_\theta(S_u)
\end{equation}
This vector $\mathbf{h}_u$ encapsulates preferences inferred strictly from historical behavior, serving as an uncontaminated foundation for the subsequent fusion process.

\subsubsection{Prompt Embedder}
To understand the prompt, we need to encode it into latent space. Hence, the textual prompt $p$ is encoded into a semantic vector $\mathbf{c}_p \in \mathbb{R}^{d}$ via a pre-trained Encoder, e.g., Sentence-BERT~\cite{reimers2019sentence}, and an MLP projector to align its dimension with $\mathbf{h}_u$. This can be formulated as:
\begin{equation}
    \mathbf{c}_p = g_{\theta}(p) = \text{MLP}(\text{Encoder(p)})
\end{equation}

\subsubsection{Signal Fusion Module}
To incorporate the prompt signal $\mathbf{c}_p$ and the user's interest representation $\mathbf{h}_u$, we propose a \textbf{Mixture-of-Experts (MoE) Tower architecture}. Since positive steering (``I want X'') and negative suppression (``Do not recommend Y'') represent different semantic operations, forcing them to share a single parameter space may lead to optimization conflicts. Therefore, we design two parallel, independent fusion blocks: a Positive Fusion Block($f_{\theta}^{+}$) and a Negative Fusion Block ($f_{\theta}^{-}$). The routing is based on the semantic indicator $c \in \{+, -\}$.
Despite their distinct objectives, both blocks share an \textbf{identical additive residual architecture} based on Multi-Head Cross-Attention (MHCA). 

For each fusion process, we utilize the user's intrinsic representation $\mathbf{h}_u$ as the Query, and the prompt vector $\mathbf{c}_p$ as both the Key and Value. The feature interaction is defined as:
\begin{equation}
    \mathbf{z}_c = \operatorname{MHCA}_c(\mathbf{Q}=\operatorname{LN}(\mathbf{h}_u), \mathbf{K}=\mathbf{c}_p, \mathbf{V}=\mathbf{c}_p)
\end{equation}
where $\operatorname{MHCA}_c$ denotes the specific attention parameters for the chosen path.
Crucially, to preserve the stability of the user's intrinsic preference, we employ a \textit{residual connection} for both paths. The prompt information is added to the original representation:
\begin{equation}
    \mathbf{h}_{\text{res}} = \mathbf{h}_u + \mathbf{z}_c
\end{equation}
Finally, the representation is processed through a path-specific Feed-Forward Network (FFN):
\begin{equation}
    \mathbf{h}_{\text{final}} = \operatorname{LN}(\mathbf{h}_{res} + \operatorname{FFN}_c(\mathbf{h}_{res}))
\end{equation}

The $\mathbf{h}_{\text{final}}$ is used as a representation that combines both historicial interest and instruction. The probability of recommending item $v \in V$ is computed as:
\begin{equation} \label{eq:final}
    P(v | \mathbf{h}_{\text{final}}) = \frac{\exp(\mathbf{h}_{final} \cdot \mathbf{e}_v)}{\sum_{v' \in \mathcal{V}} \exp(\mathbf{h}_{final} \cdot \mathbf{e}_{v'})}
\end{equation}

\subsection{Training Strategy} \label{sec:method_train}
To achieve precise controllability while maintaining robust sequential recommendation capabilities, we propose a \textbf{Unified Prediction Objective}. We formulate both positive steering and negative suppression as a likelihood maximization problem over a context-dependent \textbf{Target Item Set}, denoted as $\mathcal{V}_{\text{target}}$.

\subsubsection{Target Set Construction}
For a given training instance $(S_u, p, c)$, we are able to first construct a compliance list for user $u$, where $\mathcal{V}^u_{\text{com}} \triangleq [v \mid v \in \mathcal{V}_{\text{future}}^u \land v \text{ is compliant with } p]$, $\mathcal{V}_{\text{future}}^u$ denotes the list containing all future interacted items of user $u$. $|\mathcal{V}_{\text{com}}^u| = N$. Hereby selecting the closest $N$ samples that comply with the prompt $p$, we can control the length of the compliance list, forcing the model to focus on short-term or long-term.
The construction of $\mathcal{V}_{\text{target}}$ is dynamically adjusted to the indicator $c$:

\begin{itemize}[leftmargin=*]
    \item \textbf{Positive Steering ($c=+$):} 
    The target set $\mathcal{V}_{\text{target}}$ consists of one unique ground-truth item $v_{\text{target}}$ that is randomly sampled from the compliance list $\mathcal{V}^u_{\text{com}}$ to avoid overfitting.
    \item \textbf{Negative Suppression ($c=-$):}
    The target set $\mathcal{V}_{\text{target}}$ contains all the elements in the compliance list $\mathcal{V}^u_{\text{com}}$, denoted as $\mathcal{V}_{\text{target}} = \{ v | v \in \mathcal{V}_{\text{com}}^u\}$, to suppress the undesired items.
\end{itemize}

\subsubsection{Unified Loss Function}
A standard Cross-Entropy loss framework is adopted to optimize the model end-to-end, like Eq.~\ref{eq:final}. Here we adopt a temperature-scaled variant where:
\begin{equation}
    P(v | \mathbf{h}_{\text{final}}) = \frac{\exp(\mathbf{h}_{\text{final}} \cdot \mathbf{e}_v / \tau)}{\sum_{v' \in \mathcal{V}} \exp(\mathbf{h}_{\text{final}} \cdot \mathbf{e}_{v'} / \tau)}.
\end{equation}
Here $\mathbf{e}_v$ is the embedding of item $v$ and $\tau$ is the temperature hyperparameter.
The final training objective is defined as the cross-entropy loss averaged over the target item set $\mathcal{V}_{\text{target}}$:
\begin{equation}
\begin{aligned}
    \mathcal{L} &= - \frac{1}{|\mathcal{V}_{\text{target}}|} \sum_{v \in \mathcal{V}_{\text{target}}} \log P(v | \mathbf{h}_{final}) \\
\end{aligned}
\end{equation}

For positive steering tasks ($c=+$), the objective maximizes the likelihood of a single target item $v_{\text{target}}$. 
This forces the model to actively shift its focus beyond historical interests to satisfy the user's explicit desire while maintaining robustness, considering the random sampling when selecting $v_{\text{target}}$.
For negative suppression tasks ($c=-$), the objective operates as a multi-target optimization over the entire target set. This encourages the model to redistribute probability mass across all valid future interests, thereby implicitly suppressing the scores of restricted items through the softmax competition mechanism.

Formally, to ensure the model retains its fundamental recommendation capability while acquiring controllability, the overall training objective is formulated as a multi-task learning objective:
\begin{equation}
    \mathcal{L}_{total} = \mathcal{L}_{seq} + \mathcal{L}_{prompt}
\end{equation}
where $\mathcal{L}_{seq}$ denotes the loss for the pure sequential prediction task ($v_{t+1}$), while $\mathcal{L}_{prompt}$ corresponds to the specific instances of the loss $\mathcal{L}$ applied to the target set samples.

\subsubsection{Three-stage training}

The training of a DPR paradigm evolves through three distinct stages. \textbf{Stage 1} involves the standard pre-training of the base sequential encoder (e.g., SASRec or GRU4Rec) on the next-item prediction task to capture fundamental user behavioral patterns, following Eq. \textbf{Stage 2} serves as a baseline fine-tuning phase where the model learns to align user representations with broad category (genre) embeddings. \textbf{Stage 3}, the core contribution of this work, focuses on deep semantic alignment. Notably, the data partitioning, sequence padding, and negative sampling strategies in Stage 3 are \textbf{identical} to those in Stage 2, ensuring a controlled comparison. The primary distinction lies in the transition from coarse-grained genre prompts to fine-grained semantic prompts, supported by the following knowledge augmentation strategies. The overall training pipeline is shown in Algorithm~\ref{alg:overall}

\paragraph{Semantic Augmentation} To bridge the semantic gap between specific movie titles and abstract genres, we utilize LLMs to enrich the item metadata. For movies in the training set, we generate descriptive tags across three dimensions: \textit{Narrative} (plot and conflict), \textit{Atmosphere} (tone and style), and \textit{Appeal} (viewer motivation). This transforms coarse genre labels into explanatory semantic bridges. Such tags are utilized only in Stage 3. To rigorously test the model's generalization beyond simple keyword matching and avoid information leakage between the training and testing phases, we generate \textit{lexically distinct but semantically equivalent} variants for the testing set. LLMs are tasked with paraphrasing original tags using synonyms and alternative syntactic structures. This \textit{lexical decoupling} ensures that the model is evaluated on its understanding of latent semantics rather than surface-level patterns.

\begin{algorithm}
\caption{The Training Process of DPR Paradigm}

\label{alg:overall}
\begin{algorithmic}[1]

    \Require{Sequential Dataset $\mathcal{D}$, Target Set $\mathcal{V}_{\text{target}}$, Semantic Augmented Tags $\mathcal{T}_{aug}$}
    \Ensure{A DPR paradigm consists of a sequential encoder, a prompt embedder, and a signal fusion module.}

    \State{Stage 1: Pretrained the recommendation model $f_\theta$ given Eq.~\ref{eq:seq} on Dataset $\mathcal{D}$} 

    \State{Stage 2: Train all module except prompt embedder given Eq.~\ref{eq:final} with target set $\mathcal{V}_{\text{target}}$.} 

    \State{Stage 3: Train all module except prompt embedder given Eq.~\ref{eq:final} by replacing the genre tagets in target set $\mathcal{V}_{\text{target}}$ with semantic augmented tags $\mathcal{T}_{aug}$.}

\end{algorithmic}
\end{algorithm}

\section{Experiments}

\begin{table*}[htbp!]
\centering
\small
\caption{Main Performance comparison.}
\vspace{-10pt}
\label{tab:all}
\begin{tabular}{c|l|l|cccc|cccc}
\hline
\multirow{2}{*}{\textbf{Dataset}} & \multirow{2}{*}{\textbf{Task Type}} & \multirow{2}{*}{\textbf{Model}} & \multicolumn{4}{c|}{\textbf{SASRec}} & \multicolumn{4}{c}{\textbf{GRU4Rec}} \\
\cline{4-11}
& & & {NDCG@10} & {NDCG@20} & {Recall@10} & {Recall@20} & {NDCG@10} & {NDCG@20} & {Recall@10} & {Recall@20} \\ 
\hline
\multirow{20}{*}{\textbf{ML1M}} & \multirow{4}{*}{\textbf{Sequential}} 
& Base & 0.1728 & 0.2044 & 0.3134 & 0.4386 & 0.1569 & 0.1866 & 0.2914 & 0.4089 \\
& & \textit{+ f} & \underline{0.1800} & \underline{0.2112} & \underline{0.3235} & \underline{0.4474} & \underline{0.1816} & \underline{0.2140} & \underline{0.3214} & \underline{0.4502} \\
& & \textit{+ f + a} & 0.1410 & 0.1687 & 0.2589 & 0.3687 & 0.1261 & 0.1531 & 0.2397 & 0.3472 \\
& & DPR & \textbf{0.1838} & \textbf{0.2161} & \textbf{0.3341} & \textbf{0.4621} & \textbf{0.1817} & \textbf{0.2145} & \textbf{0.3220} & \textbf{0.4520} \\ 
\cline{2-11}
& \multirow{8}{*}{\textbf{Positive}}
& Base & 0.0781 & 0.1049 & 0.1574 & 0.2638 & 0.0673 & 0.0906 & 0.1391 & 0.2320 \\
& & \textit{+ f} & 0.0855 & 0.1108 & 0.1721 & 0.2722 & 0.0708 & 0.0954 & 0.1460 & 0.2437 \\
& & \textit{+ f + a} & 0.0812 & 0.1037 & 0.1609 & 0.2502 & 0.0748 & 0.0975 & 0.1486 & 0.2388 \\ 
\cline{3-11}
& & Filter & 0.3242 & 0.3534 & 0.5240 & 0.6388 & 0.3008 & 0.3308 & 0.4907 & 0.6088 \\
& & \textit{+ f} & \underline{0.3345} & \underline{0.3624} & \underline{0.5358} & \underline{0.6452} & \underline{0.3101} & \underline{0.3386} & \underline{0.5056} & \underline{0.6181} \\
& & \textit{+ f + a} & 0.3110 & 0.3404 & 0.5028 & 0.6184 & 0.3038 & 0.3320 & 0.4975 & 0.6085 \\ 
\cline{3-11}
& & \multirow{2}{*}{DPR} 
& \textbf{0.5748} & \textbf{0.5921} & \textbf{0.7300} & \textbf{0.7981} & \textbf{0.5501} & \textbf{0.5683} & \textbf{0.6991} & \textbf{0.7710} \\
& & & (+71.84\%) & (+63.38\%) & (+36.24\%) & (+23.70\%) & (+77.40\%) & (+67.83\%) & (+38.27\%) & (+24.74\%) \\
\cline{2-11}

& \multirow{8}{*}{\textbf{Negative}}
& Base & 0.1363 & 0.1803 & 0.1641 & 0.2653 & 0.1171 & 0.1579 & 0.1432 & 0.2375 \\
& & \textit{+ f} & {0.1463} & {0.1916} & {0.1742} & {0.2784} & 0.1230 & 0.1651 & 0.1497 & 0.2471 \\
& & \textit{+ f + a} & 0.1382 & 0.1772 & 0.1618 & 0.2507 & {0.1281} & {0.1675} & {0.1499} & {0.2403} \\  
\cline{3-11}
& & Filter & 0.1666 & 0.2147 & 0.1954 & 0.3057 & 0.1443 & 0.1884 & 0.1728 & 0.2742 \\
& & \textit{+ f} & \textbf{0.1789} & \textbf{0.2278} & \underline{0.2075} & \textbf{0.3197} & \underline{0.1509} & \underline{0.1978} & \underline{0.1792} & \underline{0.2872} \\
& & \textit{+ f + a} & 0.1599 & 0.2026 & 0.1827 & 0.2804 & 0.1465 & 0.1886 & 0.1690 & 0.2653 \\ 
\cline{3-11}
& & DPR & \begin{tabular}[c]{@{}c@{}}\underline{0.1772} \\ \small{(-0.95\%)}\end{tabular} & \begin{tabular}[c]{@{}c@{}}\underline{0.2250} \\ \small{(-1.23\%)}\end{tabular} & \begin{tabular}[c]{@{}c@{}}\textbf{0.2076} \\ \small{(+0.05\%)}\end{tabular} & \begin{tabular}[c]{@{}c@{}}\underline{0.3172} \\ \small{(-0.78\%)}\end{tabular} & \begin{tabular}[c]{@{}c@{}}\textbf{0.1741} \\ \small{(+15.37\%)}\end{tabular} & \begin{tabular}[c]{@{}c@{}}\textbf{0.2220} \\ \small{(+12.23\%)}\end{tabular} & \begin{tabular}[c]{@{}c@{}}\textbf{0.2030} \\ \small{(+13.28\%)}\end{tabular} & \begin{tabular}[c]{@{}c@{}}\textbf{0.3124} \\ \small{(+8.77\%)}\end{tabular} \\ 
\hline
\multirow{20}{*}{\textbf{MIND}} & \multirow{4}{*}{\textbf{Sequential}} 
& {Base} & 0.1058 & 0.1268 & 0.1919 & 0.2752 & 0.0977 & 0.1171 & 0.1776 & 0.2546 \\
& & \textit{+ f} & 0.1097 & 0.1310 & 0.1991 & 0.2836 & \textbf{0.1001} & \underline{0.1198} & \underline{0.1819} & \underline{0.2601} \\
& & \textit{+ f + a} & \textbf{0.1146} & \textbf{0.1380} & \textbf{0.2095} & \textbf{0.3022} & \underline{0.0997} & \textbf{0.1210} & \textbf{0.1844} & \textbf{0.2689} \\
& & {DPR} & \underline{0.1118} & \underline{0.1334} & \underline{0.2023} & \underline{0.2879} & 0.0959 & 0.1149 & 0.1745 & 0.2498 \\ 
\cline{2-11}
& \multirow{8}{*}{\textbf{Positive}}
& {Base} & 0.0405 & 0.0519 & 0.0789 & 0.1242 & 0.0378 & 0.0484 & 0.0737 & 0.1159 \\
& & \textit{+ f} & 0.0418 & 0.0539 & 0.0817 & 0.1297 & 0.0373 & 0.0477 & 0.0729 & 0.1146 \\
& & \textit{+ f + a} & 0.0591 & 0.0744 & 0.1147 & 0.1757 & 0.0498 & 0.0633 & 0.0966 & 0.1506 \\  
\cline{3-11}
& & {Filter} & 0.3199 & 0.3356 & 0.4834 & 0.5446 & 0.3122 & 0.3274 & 0.4739 & 0.5332 \\
& & \textit{+ f} & 0.3332 & 0.3491 & 0.5025 & 0.5645 & 0.3054 & 0.3204 & 0.4621 & 0.5206 \\
& & \textit{+ f + a} & \underline{0.4060} & \underline{0.4225} & \textbf{0.5964} & \textbf{0.6606} & \underline{0.3717} & \underline{0.3886} & \underline{0.5522} & \underline{0.6179} \\ 
\cline{3-11}
& & {DPR} & 
\begin{tabular}[c]{@{}c@{}}\textbf{0.4591} \\ \small{(+13.08\%)}\end{tabular} & 
\begin{tabular}[c]{@{}c@{}}\textbf{0.4751} \\ \small{(+12.45\%)}\end{tabular} & 
\begin{tabular}[c]{@{}c@{}}\underline{0.5912} \\ \small{(-0.87\%)}\end{tabular} & 
\begin{tabular}[c]{@{}c@{}}\underline{0.6543} \\ \small{(-0.95\%)}\end{tabular} & 
\begin{tabular}[c]{@{} c@ {}}\textbf{0.4860} \\ \small{(+30.75\%)}\end{tabular} & \begin{tabular}[c]{@{} c@ {}}\textbf{0.5023} \\ \small{(+29.26\%)}\end{tabular} & \begin{tabular}[c]{@{} c@ {}}\textbf{0.6237} \\ \small{(+12.95\%)}\end{tabular} & \begin{tabular}[c]{@{} c@ {}}\textbf{0.6882} \\ \small{(+11.38\%)}\end{tabular} \\
\cline{2-11}
& \multirow{8}{*}{\textbf{Negative}}
& {Base} & 0.0405 & 0.0519 & 0.0789 & 0.1242 & 0.0583 & 0.0748 & 0.0734 & 0.1158 \\
& & \textit{+ f} & 0.0418 & 0.0539 & 0.0817 & 0.1297 & 0.0572 & 0.0734 & 0.0724 & 0.1142 \\
& & \textit{+ f + a} & 0.0591 & 0.0744 & 0.1147 & 0.1757 & 0.0763 & 0.0973 & 0.0962 & 0.1501 \\  
\cline{3-11}
& & {Filter} & 0.0664 & 0.0841 & 0.0829 & 0.1284 & 0.0623 & 0.0791 & 0.0775 & 0.1207 \\
& & \textit{+ f} & 0.0691 & 0.0877 & 0.0861 & 0.1341 & 0.0613 & 0.0778 & 0.0766 & 0.1192 \\
& & \textit{+ f + a} & \textbf{0.0967} & \textbf{0.1212} & \textbf{0.1199} & \textbf{0.1827} & \underline{0.0810} & \underline{0.1023} & \underline{0.1010} & \underline{0.1559} \\ 
\cline{3-11}
& & {DPR} & \begin{tabular}[c]{@{}c@{}}\underline{0.0911} \\ \small{(-5.79\%)}\end{tabular} & \begin{tabular}[c]{@{}c@{}}\underline{0.1138} \\ \small{(-6.11\%)}\end{tabular} & \begin{tabular}[c]{@{}c@{}}\underline{0.1125} \\ \small{(-6.17\%)}\end{tabular} & \begin{tabular}[c]{@{}c@{}}\underline{0.1709} \\ \small{(-6.46\%)}\end{tabular} & \begin{tabular}[c]{@{}c@{}}\textbf{0.0868} \\ \small{(+7.16\%)}\end{tabular} & \begin{tabular}[c]{@{}c@{}}\textbf{0.1088} \\ \small{(+6.35\%)}\end{tabular} & \begin{tabular}[c]{@{}c@{}}\textbf{0.1074} \\ \small{(+6.34\%)}\end{tabular} & \begin{tabular}[c]{@{}c@{}}\textbf{0.1638} \\ \small{(+5.07\%)}\end{tabular} \\
\hline
\end{tabular}
\begin{tablenotes}
    \footnotesize \item[1] \noindent \textit{+ f} denotes fine-tuning on the NIP task (LowLR), while \textit{+ f + a} denotes implicit fine-tuning (Finetune). DPR's improvements on controllable tasks are relative to the best performing Filter baseline. Bold indicates the best performance in each category.
    \vspace{-5pt}
\end{tablenotes}
\end{table*}

In this section, we aim to evaluate and answer the following research questions:

\begin{itemize}[leftmargin=*]
    \item RQ1: How does the proposed DPR system perform compared to traditional recommendation systems and heuristic selection (based on tags)?
    \item RQ2: How does the proposed LLM-as-a-prior design compare against other LLM-as-a-posterier designs?
    \item RQ3: Is the recommendation system able to understand nuanced user requests?
\end{itemize}





\subsection{Experimental Design}

\subsubsection{Dataset Construction}

There are no existing benchmarks that are directly suitable for validating our hypothesis. Here, we modify the existing datasets, namely \textbf{MovieLens-1M} and \textbf{MIND}~\cite{wu2020mind}, to simulate shifts in user interest. Specifically, for \textbf{MIND}, we apply a 20-core filtering to focus on active users with rich interaction histories. In addition, we consider both positive and negative cases to simulate shifts in user interest. The dataset statistics are listed in Appendix~\ref{appendix:dataset}.
For each user, we partition the interaction sequence into training, validation, and test sets, using a \textbf{chronological ratio of 8:1:1} to avoid future leakage. To prevent interference from the cold-start problem during the controllable task phase, we apply strict filtering: any item that appears only in the validation or test sets and is absent from the training set is removed. This ensures that the backbone encoder has successfully learned the ID embeddings for all potential target items before the prompt-driven fine-tuning stage.
To evaluate the model's responsiveness to explicit user instructions, we construct a candidate set of future items that reflects a "heuristic interest shift." We extract $N$ (e.g., $N=3$) future items from the user's subsequent sequence as the ground truth (GT) for our controllable tasks.
The example prompts used are listed in Appendix \ref{appendix:prompts}.

\paragraph{Positive Task}
This task simulates a user making an explicit request for a new interest (e.g., ``I'm in the mood for a Comedy'').
\begin{enumerate}[topsep=0pt,noitemsep,nolistsep,leftmargin=*]
    \item Within the extracted future window of $N$ items, we identify those possessing a specific genre $g_B$ (where $g_B$ is different from the genre of $v_t$) and \textbf{randomly select one} as the ground-truth target $v_{gt}$.
    \item The model is provided with a positive prompt (e.g., ``Recommend a movie in the $g_B$ genre'') constructed based on the category of the selected target $v_{gt}$. 
\end{enumerate}

\paragraph{Negative Task}
This task simulates a user seeking to avoid a specific category while remaining open to other latent interests (e.g., ``Do not show me Horror movies'').
\begin{enumerate}[topsep=0pt,noitemsep,nolistsep,leftmargin=*]
    \item We identify the genre $g_A$ of the user's immediate next item as the category to be avoided.
    \item The \textbf{entire ground truth candidate set} of $N$ items (all with genres different from $g_A$) serves as the target for the negative request.
    \item The model receives a negative prompt (e.g., ``I don't want to watch $g_A$ movie tonight''). 
\end{enumerate}

\subsubsection{Metrics} Following the mainstream recommendation framework, we adopt the commonly-utilized metrics, namely Recall and NDCG, respectively. 

\subsubsection{Implementation Details}
We implement our DPR framework using PyTorch. Here we adopt two classic recommendation models as sequential encoders: SASRec~\cite{SASRec} and GRU4Rec~\cite{GRU4Rec}. For the embedder encoder, we adopt a pre-trained Sentence Transformer\footnote{https://huggingface.co/sentence-transformers/all-mpnet-base-v2}. For the routing indicator $c$, we utilize a zero-shot classification pipeline based on the BART-large-mnli model~\cite{lewis2020bart}, with the results shown in Appendix~\ref{appendix:intent_classification}.

\begin{table*}[!htbp]
\centering
\small
\caption{Comparison of DPR with LLM-based recommendation models on the ML1M dataset.}
\vspace{-10pt}
\label{tab:rq2_llm_comparison}
\begin{tabular}{@{}llcccccccc@{}}
\toprule
\textbf{Task} & \textbf{Metric} & \textbf{Qwen2.5-7B} & \textbf{Llama-2-7b} & \textbf{DeepSeek-V3} & \textbf{GPT-5.2} & \textbf{RecGPT-7B} & \textbf{RecLM-gen} & \textbf{DPR (SAS)} & \textbf{DPR (GRU)} \\ 
\midrule
\multirow{2}{*}{\textbf{[SEQ]}} 
& Recall@10 & 0.0296 & 0.0126 & 0.0500 & 0.0640 & 0.1864 & 0.1937 & \textbf{0.3341} & \underline{0.3220} \\
& NDCG@10   & 0.0170 & 0.0071 & 0.0227 & 0.0277 & 0.0986 & 0.1073 & \textbf{0.1838} & \underline{0.1817} \\ 
\midrule
\multirow{2}{*}{\textbf{[POS]}} 
& Recall@10 & 0.0137 & 0.0134 & 0.0263 & 0.0313 & 0.1465 & 0.3626 & \textbf{0.7300} & \underline{0.6991} \\
& NDCG@10   & 0.0078 & 0.0072 & 0.0129 & 0.0149 & 0.0745 & 0.1774 & \textbf{0.5748} & \underline{0.5501} \\ 
\midrule
\multirow{2}{*}{\textbf{[NEG]}} 
& Recall@10 & 0.0166 & 0.0103 & 0.0350 & 0.0379 & 0.1277 & 0.1163 & \textbf{0.2076} & \underline{0.2030} \\
& NDCG@10   & 0.0130 & 0.0081 & 0.0201 & 0.0228 & 0.0870 & 0.0773 & \textbf{0.1772} & \underline{0.1741} \\
\bottomrule
\end{tabular}
\begin{tablenotes}
    \footnotesize \item[1] \noindent All results are reported at Top-10. Bold indicates the best performance, and underline indicates the second best. For DeepSeek-V3 and GPT-5.2, we evaluate on a subset of 1,000 randomly sampled users.
    \vspace{-5pt}
\end{tablenotes}
\end{table*}

\subsection{Comparison with Sequential Recommendation (RQ1)}
In this section, we will compare how our promptable recommendation helps better capture users' interest against a traditional recommendation model. 



\subsubsection{Main Results and Analysis}
We now present the main experimental results. The performance on controllable tasks (Positive and Negative Prompts) is reported as the average of metrics over $N=3, 5, 10$ ground truth sets to provide a comprehensive view. For our DPR model, we report the improvement over the strong \texttt{Filter} baseline families in both relative terms. Based on the results shown in Table~\ref{tab:all}, we can make the following observations.

First, we can observe that DPR consistently achieves performance comparable to, and in some cases superior to, fine-tuned baselines (denoted as $+f$ and $+f+a$). This suggests that the introduction of DPR does not compromise the
model’s ability to capture user interests.
Second, for \textit{Positive Steering}, DPR demonstrates dominant performance across both datasets and backbones. On ML-1M, DPR achieves a remarkable \textbf{71.84\%} relative improvement in NDCG@10 over the strongest filter baseline using SASRec. Similarly, on MIND, improvements range from \textbf{13\% to 30\%}. Both observations confirm that the fusion module effectively follows the positive instruction and accurately retrieves the target item from the candidate pool.
Third, for \textit{Negative Suppression}, the results are more nuanced. With the GRU4Rec backbone, DPR consistently outperforms filter-based methods on both datasets (+15.37\% on ML-1M, +7.16\% on MIND). However, with the SASRec backbone, DPR slightly trails the strongest filter baselines (e.g., Filter $+f$ on ML-1M and Filter $+f+a$ on MIND). This suggests that while DPR learns effective soft suppression, ``hard filtering'' (removing genres from a retrieval list) remains a strong baseline under certain cases.

\paragraph{Advantage of End-to-End Ranking vs. Heuristic Filtering.}
An interesting and critical observation lies in the disparity between ranking metrics (NDCG) and retrieval metrics (Recall). Across all positive tasks, DPR's relative gain in NDCG significantly exceeds its gain in Recall. For instance, on ML-1M (SASRec), the NDCG@10 improvement is 71.84\%, whereas the Recall@10 improvement is 36.24\%. Furthermore, the improvements at the top-10 cutoff are generally more pronounced than at the top-20. 
Both observations highlight one advantage of our DPR approach. Specifically, the Filter baselines cannot alter the backbone's original ranking order. If a target item is initially ranked low (e.g., 50th) by the backbone, simply filtering out non-compliant items might only promote it marginally (e.g., to 40th). In contrast, DPR will actively alter the ranking order. The prompt vector modifies the user representation to explicitly align with the target item's semantics. This allows DPR to assign high confidence scores to the target items, ``pulling'' them to the very top of the recommendation list. This capability to re-rank based on intent, rather than just filter based on constraints, is the core driver of DPR's superior performance.

\subsection{Comparison with LLM-as-a-posterier (RQ2)}

\subsubsection{LLM-based Recommendation}

In this section, we compare the promptable recommendation to the LLM-powered recommendation, which takes the user's history as part of the prompts. 

For the LLM-based baselines, including Qwen2.5-7B, Llama-2-7b, and RecGPT-7B, we utilize their off-the-shelf pre-trained checkpoints without further fine-tuning to evaluate their zero-shot or inherent recommendation capabilities. Following the inference protocol in RecGPT~\cite{RecGPT}, we employ a beam search strategy (beam size = 10) to generate multiple response outputs. Since generative LLMs may produce diverse textual descriptions, we apply a semantic-similarity-based approach: the generated outputs are encoded into dense vectors and matched against the entire item candidate pool using FAISS to retrieve the Top-10 most similar items based on cosine similarity.
For the RecLM-gen baseline, we adapt its instruction-tuning framework to our specific data environment, covering the full suite of $I_0$-$I_3$ tasks as defined in the original work. The model is fine-tuned using the same 8:1:1 data partitioning and the same SASRec teacher model for label augmentation, ensuring a consistent knowledge base. During inference, RecLM-gen is integrated into our unified evaluation pipeline, where its performance is measured against our instruction-driven ground truth selection protocol to assess its effectiveness in capturing specific user intentions within the interaction history.


\begin{table*}[!htbp]
\caption{Comparison of DPR with LLM-based reranking models on the ML1M dataset.}
\vspace{-10pt}
\label{tab:rq2_llm_rerank}
\centering
\begin{tabular}{cc|cc|ccc|c|cc|ccc|c}
\hline
    \multicolumn{2}{c|}{Rec Model} & \multicolumn{6}{c|}{\textbf{SASRec}} & \multicolumn{6}{c}{\textbf{GRU4Rec}} \\
\hline
    \multicolumn{2}{c|}{Method} & \multicolumn{2}{c|}{Llama2-7b} & \multicolumn{3}{c|}{DeepSeek-V3} & DPR & \multicolumn{2}{c|}{Llama2-7b} & \multicolumn{3}{c}{DeepSeek-V3} & DPR \\
\hline
    \multicolumn{2}{c|}{Context} & 20 & 30 & 20 & 50 & 100 & - & 20 & 30 & 20 & 50 & 100 & - \\
\hline
    \multirow{4}{*}{\textbf{[POS]}} 
    & NDCG@10   & 0.0824 & 0.0766 & 0.2453 & 0.3950 & \underline{0.5062} & \textbf{0.5748} & 0.0700 & 0.0662 & 0.2166 & 0.3695 & \underline{0.4814} & \textbf{0.5501} \\
    & NDCG@20   & 0.1046 & 0.1013 & 0.2455 & 0.3956 & \underline{0.5076} & \textbf{0.5921} & 0.0896 & 0.0882 & 0.2167 & 0.3699 & \underline{0.4827} & \textbf{0.5683} \\
    & Recall@10 & 0.1650 & 0.1532 & 0.2530 & 0.4134 & \underline{0.5378} & \textbf{0.7300} & 0.1447 & 0.1343 & 0.2226 & 0.3861 & \underline{0.5107} & \textbf{0.6991} \\
    & Recall@20 & 0.2537 & 0.2516 & 0.2537 & 0.4158 & \underline{0.5429} & \textbf{0.7981} & 0.2233 & 0.2223 & 0.2233 & 0.3879 & \underline{0.5157} & \textbf{0.7710} \\
\hline
    \multirow{4}{*}{\textbf{[NEG]}}  
    & NDCG@10   & \underline{0.1215} & 0.1136 & 0.1083 & 0.0876 & 0.0795 & \textbf{0.1772} & \underline{0.1050} & 0.0985 & 0.0938 & 0.0781 & 0.0722 & \textbf{0.1741} \\
    & Recall@10 & \underline{0.1448} & 0.1361 & 0.1344 & 0.1008 & 0.0901 & \textbf{0.2250} & \underline{0.1272} & 0.1201 & 0.1186 & 0.0927 & 0.0842 & \textbf{0.2220} \\
    & NDCG@20   & \underline{0.1694} & 0.1628 & 0.1602 & 0.1261 & 0.1081 & \textbf{0.2076} & \underline{0.1486} & 0.1429 & 0.1408 & 0.1138 & 0.0984 & \textbf{0.2030} \\
    & Recall@20 & \underline{0.2552} & 0.2502 & \underline{0.2552} & 0.1901 & 0.1561 & \textbf{0.3172} & \underline{0.2284} & 0.2241 & \underline{0.2284} & 0.1762 & 0.1450 & \textbf{0.3124} \\
\hline
\end{tabular}
\begin{tablenotes}
    \footnotesize \item[1] \noindent Context indicates how many of the top-K recommendation result from the recommendation method is included in the LLM context.
    \vspace{-5pt}
\end{tablenotes}
\end{table*}

\textbf{Discussion on RQ2: Comparison with LLM-based Models.} 
As shown in Table~\ref{tab:rq2_llm_comparison}, two key observations emerge from the comparison on the ML1M dataset:
First, general-purpose LLMs (e.g., Qwen2.5, Llama-2, DeepSeek-V3) exhibit limited zero-shot performance, indicating that inherent semantic knowledge alone is insufficient for precise item ranking. In contrast, domain-specific models (RecGPT-7B, RecLM-gen) achieve significantly higher performance across all metrics. This validates that encoding collaborative information from recommendation data into LLM remains essential for aligning LLMs with user interaction patterns.
Second, despite the improvements in fine-tuned LLMs, our DPR framework consistently outperforms the best LLM baseline by a substantial margin across all tasks. For example, in the [POS] task, DPR (SAS) achieves a Recall@10 of 0.7300 compared to 0.3626 for RecLM-gen. This demonstrates that while generative LLMs can understand instructions, our specialized decoupled fusion architecture is more effective in strictly steering latent user representations to satisfy specific positive and negative constraints without compromising the integrity of sequential modeling.

\subsubsection{LLM-based Reranking}
We further investigate a two-stage \textit{retrieve-and-rerank} pipeline, where LLMs (Llama2-7b, DeepSeek-V3) rerank the top-$K$ candidates retrieved by the backbone model. Table~\ref{tab:rq2_llm_rerank} presents the performance across varying candidate sizes ($K \in \{20, 30, 50, 100\}$). We can observe the followings:

First, we observe divergent scaling behaviors in positive steering. In the [POS] task, traditional LLMs like Llama2-7b exhibit a performance decline as the candidate set grows, due to information overload. In sharp contrast, DeepSeek-V3 demonstrates a positive correlation between context size and accuracy. Expanding $K$ from 20 to 100 significantly boosts its NDCG@10 (e.g., $0.2453 \to 0.5062$ with SASRec), indicating LLM's superior capability to mine latent positive targets from larger candidate pools.
Second, we observe universal vulnerability in negative suppression task. For the [NEG] task, all LLM rerankers degrade in performance as $K$ increases. This reveals a fundamental limitation: LLMs are highly sensitive to noise when executing ``exclusion'' logic. As the candidate pool expands, the complexity of filtering out prohibited items increases, leading models to inadvertently rank restricted items higher.
Finally, we observe that despite DeepSeek-V3's impressive scaling in positive tasks, our DPR framework still outperforms the best reranking configuration (0.5748 vs. 0.5062 in POS NDCG@10). Crucially, DPR achieves this robust control in a single end-to-end inference step, avoiding the high latency and noise sensitivity inherent in large-context LLM reranking.

\subsection{Capture of Implicit User Intent (RQ3)}

In this section, we move beyond explicit genre constraints to investigate the model's ability to interpret nuanced natural-language requests. We employ an LLM-based simulation framework where virtual users powered by external LLM express complex, implicit needs, and an LLM judge evaluates the relevance of recommendations.

\subsubsection{Evaluation Protocol}
\paragraph{User Simulation}
To generate realistic and diverse user queries, we utilize DeepSeek-V3 as a simulator. Given a user's interaction history, the simulator is prompted to articulate a current viewing desire in natural language without explicitly naming genres (see Appendix~\ref{appendix:prompts}). This ensures the request reflects high-level semantics such as \textit{''I want something chaotic to distract me''} rather than simple keyword matching.

\paragraph{LLM-as-a-Judge}
Evaluating subjective alignment is challenging for traditional metrics. We adopt an \textit{LLM-as-a-Judge} paradigm, which is conceptually aligned with prior work aiming to evaluates LLM's inherent recommendation capabilities~\cite{he2023crs}. Here a separate LLM instance scores the recommended list (top-10) on a 1--10 scale across three dimensions: \textbf{History Alignment} (consistency with long-term taste), \textbf{Intent Fulfillment} (adherence to the current specific request), and an \textbf{Overall Score}. The detailed judging criteria are provided in Appendix~\ref{appendix:prompts}.

\subsubsection{Results and Analysis}
Table~\ref{tab:rq3_judge_scores} reports the comparative performance of DPR against baselines in satisfying implicit user intents.
The results demonstrate that DPR effectively bridges the gap between natural language instruction and latent item semantics. DPR achieves the highest \textbf{Overall Score}, outperforming all baselines. This confirms that our decoupled architecture successfully balances the dual objectives: satisfying the immediate crave (the prompt) while remaining grounded in the user's established preferences (the history).
This confirms that the semantic alignment learned during our fine-tuning stage generalizes well to unstructured, conversational scenarios, empowering the model to handle the ambiguity inherent in real-world user interactions.

\begin{table}[!htbp]
\centering
\caption{LLM-as-a-Judge evaluation results for nuanced semantic requests on ML1M. Scores range from 1 to 10. \texttt{DPR-Rand} serves as an ablation to test prompt sensitivity by providing mismatched prompts.}
\vspace{-10pt}
\label{tab:rq3_judge_scores}
\resizebox{0.98\columnwidth}{!}{
\begin{tabular}{lccc}
\toprule
\multirow{2}{*}{\textbf{Model}} & \textbf{Intent} & \textbf{History} & \textbf{Overall} \\ 
& \textbf{Fulfillment} & \textbf{Alignment} & \textbf{Score} \\
\midrule
\texttt{SASRec} & 4.076 & \textbf{6.090} & 4.708 \\
\texttt{GPT-5.2} & \textbf{6.231} & 5.007 & \underline{5.619} \\
\texttt{DPR-Rand (Ablation)} & 3.288 & 4.816 & 3.684 \\
\texttt{DPR (Ours)} & \underline{5.888} & \underline{5.918} & \textbf{5.932} \\
\bottomrule
\end{tabular}
}
\vspace{-10pt}
\end{table}






\section{Ablation Study}

\subsection{Ablation on Stage Design}
\label{sec:ablation_stage}

To validate the necessity of the three-stage training framework proposed in Section~\ref{sec:methodology}, specifically the inclusion of the coarse-grained genre alignment (Stage 2), we conducted an ablation study on both ML-1M and MIND datasets. We compare the full {3-Stage (3-S)} framework against a {2-Stage (2-S)} variant. In the 2-S setting, the model skips the genre-based alignment and fine-tunes the pre-trained backbone directly using the fine-grained semantic tags generated by LLMs. The results are reported in Table~\ref{tab:stage_ablation}.

As observed, the 3-S strategy consistently outperforms the 2-S variant across all metrics and backbones. Removing the intermediate genre alignment stage results in performance degradation. For instance, on the ML-1M dataset with the SASRec backbone, the NDCG@10 for positive steering ([POS]) and negative suppression ([NEG]) drops by 7.85\% and 13.26\%, respectively. This empirical evidence suggests that directly mapping user representations to a complex, fine-grained semantic space is optimizationally difficult. The coarse-grained alignment in Stage 2 effectively serves as a scaffold, establishing a stable semantic foundation that facilitates the subsequent deep semantic alignment in Stage 3.

\subsection{Ablation on Loss Design}
\label{sec:ablation_loss}
To investigate the contribution of the specific terms in Eq.~\ref{eq:final}, we conduct an ablation study by isolating the control-specific loss terms. We compare the complete model (\textbf{Full}) against two variants: (1) \textbf{Pos}, which is trained without negative targets; and (2) \textbf{Neg}, which is trained without positive targets. The results are summarized in Table~\ref{tab:loss_ablation}. We can make the following observations.

First, the results demonstrate that both $\mathcal{L}_{pos}$ and $\mathcal{L}_{neg}$ are critical for their respective control capabilities. 
On the positive steering task ({[POS]}), the \textit{Neg} variant exhibits a catastrophic performance collapse (e.g., N@10 drops from 0.5748 to 0.1670 on ML-1M with SASRec). This indicates that the ability to retrieve specific items based on semantic prompts cannot be implicitly acquired through suppression training or pure sequential modeling; explicit positive alignment is indispensable.
Similarly, on the negative suppression task ({[NEG]}), the \textit{Pos} variant consistently underperforms the \textit{Full} model (e.g., N@10 decreases from 0.1772 to 0.1634 on ML-1M). This confirms that our proposed implicit suppression strategy—optimizing over the constrained candidate set $\mathcal{S}_{cand}$—is essential for teaching the model to effectively filter out prohibited items.
Additionally, the \textit{Full} model achieves balanced performance comparable to the specialized single-task variants. For instance, on [NEG] tasks, the \textit{Full} model matches the performance of the \textit{Neg} variant (0.1772 vs. 0.1772 on ML-1M). While there is a slight trade-off in some positive metrics compared to the \textit{Pos} variant (e.g., on MIND), the \textit{Full} model maintains robust controllability across both directions. This suggests that our decoupled architecture and unified objective successfully facilitate multi-task learning, enabling the model to master both directional steering and constraint suppression simultaneously without significant destructive interference.

\subsection{Ablation on Tower Architecture Design}
\label{sec:ablation_tower}
To justify the architectural decision of using distinct fusion pathways for conflicting control signals, we evaluate the impact of the decoupling strategy. We compare our proposed \textbf{Two-Tower (2-T)} architecture—where positive and negative prompts are processed by separate stacks of fusion blocks—against a \textbf{Single-Tower (1-T)} variant. In the 1-T variant, a shared set of MHCA parameters handles both positive steering and negative suppression tasks simultaneously. The results are presented in Table~\ref{tab:tower_ablation}.

First, we observe substantial performance degradation when using the Single-Tower architecture across all metrics. For example, on the ML-1M dataset using SASRec, the NDCG@10 decreases by \textbf{27.40\%} for the positive task and by \textbf{35.33\%} for the negative task. Similarly, on the MIND dataset, the performance drop for the negative task exceeds \textbf{40\%}.
This significant gap validates our hypothesis that positive steering and negative suppression are fundamentally divergent operations. Positive steering requires the model to \textit{enhance} semantic features related to the prompt (feature injection), whereas negative suppression requires the model to \textit{neutralize} specific features (feature rejection). Forcing a single set of parameters to learn these opposing transformations simultaneously leads to severe optimization conflicts and parameter interference. The Two-Tower architecture effectively resolves this by allocating independent parameter spaces for each control type, allowing the model to specialize in both feature excitation and inhibition without conflicting each other.
\section{Conclusion}

In this work, we discuss a simple yet unfulfilled request: instructing recommendation models throughout natural language. After identifying the drawbacks of existing paradigms, we propose the Decoupled Promptable Sequential Recommendation (DPR) framework, a model-agnostic paradigm that empowers conventional backbones to natively follow natural-language prompts. DPR outperforms existing baselines under various scenarios. Ablation study further demonstrates the necessity of various designs.


\bibliographystyle{ACM-Reference-Format}
\bibliography{reference}

\appendix
\section{Datasets and Preprocessing}
\label{appendix:dataset}

\begin{table}[h!]
\centering
\caption{Statistics of datasets.}
\label{tab:dataset_stats}
\small
    \resizebox{0.98\columnwidth}{!}{
\begin{tabular}{lrrrrr}
    \hline
    \textbf{Dataset} & \textbf{\#Users} & \textbf{\#Items} & \textbf{\#Int.} & \textbf{Avg. Len.} & \textbf{Sparsity} \\
    \hline
    ML-1M   & 6{,}040   & 3{,}667    & 1{,}000{,}155    & 165.59 & 95.48\% \\
    MIND    & 265{,}016 & 21{,}901   & 12{,}376{,}034   & 46.70  & 99.79\% \\
    \hline
\end{tabular}}
\end{table}

\section{Experimental Prompt Templates}
\label{appendix:prompts}

This appendix details the prompts we utilized.

\subsection{Experimental Prompt Templates}
To simulate explicit user instructions, we construct prompts by injecting target or avoided attributes into predefined templates, such as \textit{``I feel like watching a \{category\} film. What should I see?''} for positive steering, and \textit{``I want to watch something different, so avoid the \{category\} type.''} for negative suppression. {Specifically, we curated a pool of 20 distinct templates for each task to ensure linguistic diversity.} Crucially, to rigorously evaluate the model's semantic generalization capabilities, the template pools used during training and evaluation are strictly disjoint. This design ensures that the fusion module learns to interpret underlying user intents rather than merely memorizing specific lexical patterns.

\subsection{Training Tag Generation Prompt (Stage 3)}

To illustrate the output of the semantic knowledge augmentation process, consider the movie \textit{Toy Story} (ID: 1), which is originally associated with the broad genre \textit{``Animation''}. Our framework generates the following three multi-dimensional tags to bridge the semantic gap: 
(i) \textit{``Toys come to life with secret adventures''} (Narrative focus); 
(ii) \textit{``Whimsical and heartwarming childhood nostalgia''} (Atmosphere focus); 
(iii) \textit{``Family-friendly imaginative storytelling with emotional depth''} (Appeal focus). 

This transformed metadata allows the model to learn fine-grained semantic alignment beyond coarse categories. The specific prompt used for this generation process is provided below:

\begin{lstlisting}[breaklines=true, basicstyle=\small\ttfamily]
[System Prompt]
You are a film expert specializing in semantic tagging for recommendation systems. Your task is to bridge the gap between specific Movie Titles and their broad Genres.

[User Prompt]
Here are {N} movies with their Titles and assigned Genres:
{items_block}

**Task:**
For EACH movie, generate **3 Descriptive Semantic Tags**.

**Requirements for Tags:**
1. **Bridging Logic:** The tags must explain *why* this specific Title belongs to these Genres. Connect the specific content/plot/vibe to the abstract genre category.
2. **Diverse Angles:**
   - Tag 1: Focus on **Narrative/Plot** (What happens?).
   - Tag 2: Focus on **Atmosphere/Tone** (How does it feel?).
   - Tag 3: Focus on **Viewer Experience/Appeal** (Why do people watch it?).
3. **Constraint:** Do NOT strictly repeat the Genre name. Use descriptive words like "high-octane combat", "explosive set pieces", etc.
4. **Length:** Keep each tag concise (5-12 words).

**Output Format (JSON):**
[
    {
        "id": "1", 
        "tags": [
            "Toys come to life with secret adventures", 
            "Whimsical and heartwarming childhood nostalgia", 
            "Family-friendly imaginative storytelling with emotional depth"
        ]
    },
    ...
]
\end{lstlisting}

\subsection{Evaluation Tag Rewriting (Stage 3)}

To rigorously evaluate the model's ability to understand latent semantics rather than relying on surface-level keyword matching, we generate lexically distinct but semantically equivalent variants for the testing set. Using the movie \textit{Toy Story} (ID: 1) as an example, the original training tags are rephrased as follows:
(i) \textit{``Playthings awaken to embark on hidden escapades''} (Narrative); 
(ii) \textit{``Enchanting and tender evocation of youthful memories''} (Atmosphere); 
(iii) \textit{``Inventive tale for all ages, rich with heartfelt layers''} (Experience). 

By comparing these with the training tags (e.g., \textit{``Toys''} $\to$ \textit{``Playthings''}, \textit{``Nostalgia''} $\to$ \textit{``Youthful memories''}), we ensure that the model must generalize across vocabularies to satisfy user intents. The prompt used for this paraphrasing task is detailed below:

\begin{lstlisting}[breaklines=true, basicstyle=\small\ttfamily]
[System Prompt]
You are an expert editor and semantic linguist. Your task is to **rewrite** movie descriptions to be **lexically distinct** (using different words) but **semantically equivalent** (keeping the exact same meaning and depth).

[User Prompt]
Here are {N} movies. Each comes with its Title, Genres, and 3 specific "Original Description" tags.
{items_block}

**Task:**
For EACH movie, rewrite the 3 "Original Description" tags into 3 "New Evaluation" tags.

**Strict Requirements:**
1. **Paraphrase, Don't Summarize:** The new tags must maintain the **same level of detail and length** as the original. Do NOT make them shorter.
2. **Vocabulary Shift:** Use synonyms, different sentence structures, and alternative adjectives. 
   - Example: "Dystopian" -> "Oppressive futuristic society"; "Heartwarming" -> "Emotionally uplifting".
3. **Accuracy:** Do not change the facts or the vibe. Just change the wording.
4. **Structure:** 
   - New Tag 1 corresponds to Original Tag 1 (Narrative).
   - New Tag 2 corresponds to Original Tag 2 (Atmosphere).
   - New Tag 3 corresponds to Original Tag 3 (Experience).

**Output Format (JSON):**
[
    {
        "id": "1", 
        "tags": [
            "Playthings awaken to embark on hidden escapades", 
            "Enchanting and tender evocation of youthful memories", 
            "Inventive tale for all ages, rich with heartfelt layers"
        ]
    },
    ...
]
\end{lstlisting}

\subsection{User Simulation Prompt (RQ3)}
\begin{lstlisting}
[System Prompt]
You are an AI simulating a movie enthusiast with a rich viewing history. 
Your goal is to express what you want to watch *right now* using natural, human-like language. 
You never ask for genres directly; instead, you express feelings, vibes, specific plot cravings, or viewing occasions.

[User Prompt]
Here is the user's movie watch history (Last 20 items):
<history>
{user_history_str}
</history>

**Task:**
Generate **THREE** distinct requests representing three different "Search Modes" for this user. 

**Mode 1: The Mood/Vibe Seeker**
- Express an emotional state or a desired atmosphere.
- Example: "I've had a stressful week, I need something chaotic to distract me," or "I want a movie that feels like a rainy Sunday afternoon."

**Mode 2: The Specific Plot Craving**
- Describe a specific narrative hook or character dynamic without naming the genre.
- Example: "Show me a story about a heist that goes wrong," or "I want to watch someone struggle against isolation in nature."

**Mode 3: The Occasion/Abstract User**
- Describe a viewing context or use an abstract metaphor.
- Example: "It's date night, need something not too heavy," or "I want a movie that tastes like expensive wine."

**Constraints:**
1. **Strictly NO explicit genre names** (e.g., ban "Horror", "Action", "Romance").
2. Keep it conversational and natural (1st person).
3. Ensure the three requests are distinct from each other.

**Output Format (JSON):**
[
  {
    "implied_intent": "[The underlying genre/theme you are hinting at]",
    "statement": "Your generated natural language request here."
  },
  ...
]
\end{lstlisting}

\subsection{LLM Judge Prompt (RQ3)}
\begin{lstlisting}
[System Prompt]
You are a sophisticated movie recommendation critic. You evaluate how well a list of movies satisfies a user's complex, natural language request.
You understand that users often describe *vibes* or *occasions* rather than technical movie attributes.

[User Prompt]
**Context:**
1. **User History (Long-term Taste):**
<history>
{user_history_str}
</history>

2. **User Request (Current Intent):**
<request>
"{user_request}"
</request>

**System Recommendations:**
<recommendations>
{recommendations_with_info}
</recommendations>

**Evaluation Task:**
Rate the recommendation list by providing three separate scores from 1 (worst) to 10 (best).

**1. History Alignment Score:**
Does this list look like something this user would generally enjoy, ignoring the current specific request?
- 9-10: Perfect fit for their long-term habits.
- 1-2: Alien to their established taste.

**2. Intent Fulfillment Score:**
**Crucial:** How well does the list capture the **spirit, mood, or specific scenario** described in the request?
- If the user asked for "stress relief," are these movies distracting/fun?
- If the user asked for "isolation," do these movies convey loneliness?
- Do NOT look for exact keyword matches. Look for **Semantic Alignment**.
- 9-10: Perfectly captures the requested vibe/plot.
- 1-2: Completely misses the point of the user's words.

**3. Overall Score:**
A holistic score combining history compatibility and intent satisfaction.
- A good recommendation must solve the user's *current* problem (the request) without violating their *fundamental* taste (history).
- Weight **Intent Fulfillment** heavily.

**Output Format:**
Provide the output strictly in the following JSON format (do not include markdown fencing).

{
  "scores": {
    "history_alignment": <int 1-10>,
    "intent_fulfillment": <int 1-10>,
    "overall_score": <int 1-10>
  },
  "reasoning": {
    "history_alignment": "<Brief justification>",
    "intent_fulfillment": "<Brief justification, focusing on the vibe/semantic match>",
    "overall_score": "<Final verdict>"
  }
}
\end{lstlisting}

\subsection{Movie Summary Generation Prompt}
\begin{lstlisting}
[System Prompt]
You are a knowledgeable film critic. Provide accurate movie summaries.

[User Prompt]
You are tasked with generating a summary of the movie: "{movie_title}".
This summary should be maximally helpful for deciding if it fits a specific mood or vibe.

Include:
1. Genre and Tone (e.g., dark, lighthearted, tense, slow-burn).
2. Main Themes.

**Constraint:** Keep it concise (max 40-50 words). No spoilers.
\end{lstlisting}

\section{Automatic Intent Classification for Routing}
\label{appendix:intent_classification}

To operationalize the control indicator $c$ without manual labeling or specialized training, we employ a zero-shot classification approach based on the \texttt{BART-large-mnli} model~\cite{lewis2020bart}. This approach leverages the Natural Language Inference (NLI) capability of the model to determine the logical relationship between a user prompt and a predefined intent.

\subsection{Classification Protocol}
We define the intent detection as a binary classification task with two candidate labels: ``positive preference'' (corresponding to $c = +$) and ``negative preference'' (corresponding to $c = -$). 

To enhance the model's logical reasoning, we utilize a \textit{hypothesis template}: \textbf{``This text expresses a \{\}.''} During inference, the BART model treats the user prompt as the ``premise'' and the filled template (e.g., ``This text expresses a positive preference.'') as the ``hypothesis.'' The model then calculates the probability of entailment between the premise and each hypothesis. The label with the highest entailment score is selected as the predicted intent.

\subsection{Data Pre-processing and Verification}
During the system's development and verification, we processed the prompt templates (e.g., \textit{``I am in the mood for \{category\}''}) by replacing placeholders with specific concrete instances (such as ``sports'') to ensure the model could process the logic in a contextualized manner. 

We conducted a rigorous evaluation using the full set of training and evaluation prompts. The BART-based classifier achieved a \textbf{100\% accuracy rate} across all samples. This perfect classification performance ensures that the routing mechanism within our dual-path architecture is virtually flawless, providing a reliable foundation for the subsequent decoupled fusion process. By using this zero-shot approach, we demonstrate that user intent routing can be achieved efficiently without additional annotated data or intent-specific training.

\section{Additional Experiment}

\subsection{Detailed results on various $N$ settings}
Here, we report the individual results for the $N=3, 5, 10$ settings in Table~\ref{tab:all}. Specifically, we have Table~\ref{tab:ml1m_results}, listing results for $N=3,5,10$ on the MovieLens-1M dataset, and Table~\ref{tab:mind_topk}, listing results for $N=3,5$ on the MIND dataset. Notice that the MIND dataset lacks the $N=10$ setting due to its sequence length.

\begin{table*}[htbp!]
\centering
\small
\caption{Performance comparison on the ML-1M dataset with varied $N$ settings.}
\vspace{-10pt}
\label{tab:ml1m_results}
\begin{tabular}{@{}llcccc|cccc@{}}
\hline
\multirow{2}{*}{\textbf{Task Type}} & \multirow{2}{*}{\textbf{Model}} & \multicolumn{4}{c|}{\textbf{SASRec}} & \multicolumn{4}{c}{\textbf{GRU4Rec}} \\
\cline{3-10}
& & {NDCG@10} & {NDCG@20} & {Recall@10} & {Recall@20} & {NDCG@10} & {NDCG@20} & {Recall@10} & {Recall@20} \\ 
\hline
\multirow{7}{*}{\begin{tabular}[c]{@{}l@{}}\textbf{Positive}\\ (\textbf{N = 3})\end{tabular}} 
& {Base} & 0.0992 & 0.1293 & 0.1978 & 0.3171 & 0.0873 & 0.1155 & 0.1780 & 0.2898 \\
& \textit{+ f} & 0.1084 & 0.1373 & 0.2169 & 0.3320 & 0.0915 & 0.1204 & 0.1885 & 0.3034 \\
& \textit{+ f + a} & 0.0990 & 0.1237 & 0.1930 & 0.2912 & 0.0915 & 0.1155 & 0.1784 & 0.2737 \\
\cline{2-10}
& {Filter} & 0.3705 & 0.3989 & 0.5781 & 0.6896 & 0.3496 & 0.3794 & 0.5495 & 0.6674 \\
& \textit{+ f} & 0.3840 & 0.4117 & 0.5913 & 0.7001 & 0.3568 & 0.3852 & 0.5625 & 0.6746 \\
& \textit{+ f + a} & 0.3466 & 0.3759 & 0.5425 & 0.6575 & 0.3377 & 0.3661 & 0.5365 & 0.6485 \\
\cline{2-10}
& {DPR} & 
\begin{tabular}[c]{@{}c@{}}\textbf{0.6159} \\ \small{(+60.39\%)}\end{tabular} & 
\begin{tabular}[c]{@{}c@{}}\textbf{0.6315} \\ \small{(+53.39\%)}\end{tabular} & 
\begin{tabular}[c]{@{}c@{}}\textbf{0.7664} \\ \small{(+29.61\%)}\end{tabular} & 
\begin{tabular}[c]{@{}c@{}}\textbf{0.8279} \\ \small{(+18.25\%)}\end{tabular} & 
\begin{tabular}[c]{@{} c@ {}}\textbf{0.5929} \\ \small{(+66.17\%)}\end{tabular} & 
\begin{tabular}[c]{@{} c@ {}}\textbf{0.6097} \\ \small{(+58.28\%)}\end{tabular} & 
\begin{tabular}[c]{@{} c@ {}}\textbf{0.7410} \\ \small{(+31.73\%)}\end{tabular} & 
\begin{tabular}[c]{@{} c@ {}}\textbf{0.8071} \\ \small{(+19.64\%)}\end{tabular} \\
\hline
\multirow{7}{*}{\begin{tabular}[c]{@{}l@{}}\textbf{Negative}\\ (\textbf{N = 3})\end{tabular}} 
& {Base} & 0.1402 & 0.1809 & 0.2005 & 0.3154 & 0.1223 & 0.1609 & 0.1786 & 0.2874 \\
& \textit{+ f} & 0.1505 & 0.1922 & 0.2136 & 0.3313 & 0.1275 & 0.1673 & 0.1858 & 0.2983 \\
& \textit{+ f + a} & 0.1360 & 0.1707 & 0.1909 & 0.2887 & 0.1270 & 0.1623 & 0.1779 & 0.2779 \\
\cline{2-10}
& {Filter} & 0.1717 & 0.2152 & 0.2380 & 0.3606 & 0.1517 & 0.1921 & 0.2159 & 0.3299 \\
& \textit{+ f} & \textbf{0.1850} & \textbf{0.2290} & \textbf{0.2537} & \textbf{0.3779} & 0.1570 & 0.2002 & 0.2219 & 0.3436 \\
& \textit{+ f + a} & 0.1571 & 0.1948 & 0.2142 & 0.3206 & 0.1451 & 0.1829 & 0.1989 & 0.3058 \\
\cline{2-10}
& {DPR} & 
\begin{tabular}[c]{@{}c@{}}0.1821 \\ \small{(-1.57\%)}\end{tabular} & 
\begin{tabular}[c]{@{}c@{}}0.2255 \\ \small{(-1.53\%)}\end{tabular} & 
\begin{tabular}[c]{@{}c@{}}0.2530 \\ \small{(-0.28\%)}\end{tabular} & 
\begin{tabular}[c]{@{}c@{}}0.3754 \\ \small{(-0.66\%)}\end{tabular} & 
\begin{tabular}[c]{@{}c@{}}\textbf{0.1806} \\ \small{(+15.03\%)}\end{tabular} & 
\begin{tabular}[c]{@{}c@{}}\textbf{0.2235} \\ \small{(+11.64\%)}\end{tabular} & 
\begin{tabular}[c]{@{}c@{}}\textbf{0.2504} \\ \small{(+12.84\%)}\end{tabular} & 
\begin{tabular}[c]{@{}c@{}}\textbf{0.3712} \\ \small{(+8.03\%)}\end{tabular} \\
\hline\hline
\multirow{7}{*}{\begin{tabular}[c]{@{}l@{}}\textbf{Positive}\\ (\textbf{N = 5})\end{tabular}} 
& {Base} & 0.0779 & 0.1066 & 0.1581 & 0.2721 & 0.0708 & 0.0948 & 0.1453 & 0.2406 \\
& \textit{+ f} & 0.0859 & 0.1118 & 0.1747 & 0.2776 & 0.0743 & 0.0995 & 0.1513 & 0.2513 \\
& \textit{+ f + a} & 0.0814 & 0.1043 & 0.1633 & 0.2544 & 0.0718 & 0.0951 & 0.1484 & 0.2409 \\
\cline{2-10}
& {Filter} & 0.3325 & 0.3621 & 0.5401 & 0.6562 & 0.3095 & 0.3404 & 0.5073 & 0.6286 \\
& \textit{+ f} & 0.3419 & 0.3702 & 0.5508 & 0.6625 & 0.3189 & 0.3486 & 0.5240 & 0.6414 \\
& \textit{+ f + a} & 0.3165 & 0.3445 & 0.5146 & 0.6250 & 0.3046 & 0.3347 & 0.5003 & 0.6193 \\
\cline{2-10}
& {DPR} & 
\begin{tabular}[c]{@{}c@{}}\textbf{0.5841} \\ \small{(+70.84\%)}\end{tabular} & 
\begin{tabular}[c]{@{}c@{}}\textbf{0.6014} \\ \small{(+62.45\%)}\end{tabular} & 
\begin{tabular}[c]{@{}c@{}}\textbf{0.7401} \\ \small{(+34.37\%)}\end{tabular} & 
\begin{tabular}[c]{@{}c@{}}\textbf{0.8082} \\ \small{(+21.99\%)}\end{tabular} & 
\begin{tabular}[c]{@{} c@ {}}\textbf{0.5608} \\ \small{(+75.85\%)}\end{tabular} & 
\begin{tabular}[c]{@{} c@ {}}\textbf{0.5784} \\ \small{(+65.92\%)}\end{tabular} & 
\begin{tabular}[c]{@{} c@ {}}\textbf{0.7087} \\ \small{(+35.25\%)}\end{tabular} & 
\begin{tabular}[c]{@{} c@ {}}\textbf{0.7778} \\ \small{(+21.27\%)}\end{tabular} \\
\hline
\multirow{7}{*}{\begin{tabular}[c]{@{}l@{}}\textbf{Negative}\\ (\textbf{N = 5})\end{tabular}} 
& {Base} & 0.1400 & 0.1841 & 0.1701 & 0.2734 & 0.1221 & 0.1641 & 0.1492 & 0.2476 \\
& \textit{+ f} & 0.1500 & 0.1962 & 0.1797 & 0.2880 & 0.1285 & 0.1717 & 0.1558 & 0.2571 \\
& \textit{+ f + a} & 0.1408 & 0.1788 & 0.1670 & 0.2559 & 0.1298 & 0.1687 & 0.1536 & 0.2449 \\
\cline{2-10}
& {Filter} & 0.1709 & 0.2193 & 0.2021 & 0.3155 & 0.1503 & 0.1954 & 0.1796 & 0.2852 \\
& \textit{+ f} & \textbf{0.1834} & \textbf{0.2334} & \textbf{0.2138} & \textbf{0.3311} & 0.1576 & 0.2054 & 0.1862 & 0.2985 \\
& \textit{+ f + a} & 0.1623 & 0.2043 & 0.1881 & 0.2866 & 0.1481 & 0.1895 & 0.1732 & 0.2703 \\
\cline{2-10}
& {DPR} & 
\begin{tabular}[c]{@{}c@{}}0.1826 \\ \small{(-0.44\%)}\end{tabular} & 
\begin{tabular}[c]{@{}c@{}}0.2305 \\ \small{(-1.24\%)}\end{tabular} & 
\begin{tabular}[c]{@{}c@{}}0.2160 \\ \small{(+1.03\%)}\end{tabular} & 
\begin{tabular}[c]{@{}c@{}}0.3281 \\ \small{(-0.91\%)}\end{tabular} & 
\begin{tabular}[c]{@{}c@{}}\textbf{0.1808} \\ \small{(+14.72\%)}\end{tabular} & 
\begin{tabular}[c]{@{}c@{}}\textbf{0.2292} \\ \small{(+11.59\%)}\end{tabular} & 
\begin{tabular}[c]{@{}c@{}}\textbf{0.2111} \\ \small{(+13.37\%)}\end{tabular} & 
\begin{tabular}[c]{@{}c@{}}\textbf{0.3241} \\ \small{(+8.58\%)}\end{tabular} \\
\hline\hline
\multirow{7}{*}{\begin{tabular}[c]{@{}l@{}}\textbf{Positive}\\ (\textbf{N = 10})\end{tabular}} 
& {Base} & 0.0573 & 0.0789 & 0.1164 & 0.2023 & 0.0436 & 0.0615 & 0.0939 & 0.1656 \\
& \textit{+ f} & 0.0623 & 0.0832 & 0.1245 & 0.2071 & 0.0465 & 0.0662 & 0.0983 & 0.1765 \\
& \textit{+ f + a} & 0.0633 & 0.0831 & 0.1265 & 0.2051 & 0.0611 & 0.0819 & 0.1189 & 0.2019 \\
\cline{2-10}
& {Filter} & 0.2696 & 0.2992 & 0.4537 & 0.5705 & 0.2434 & 0.2727 & 0.4154 & 0.5302 \\
& \textit{+ f} & \textbf{0.2777} & 0.3052 & 0.4654 & 0.5729 & 0.2548 & 0.2820 & 0.4303 & 0.5383 \\
& \textit{+ f + a} & 0.2700 & 0.3007 & 0.4512 & 0.5725 & 0.2691 & 0.2950 & 0.4557 & 0.5576 \\
\cline{2-10}
& {DPR} & 
\begin{tabular}[c]{@{}c@{}}\textbf{0.5245} \\ \small{(+88.87\%)}\end{tabular} & 
\begin{tabular}[c]{@{}c@{}}\textbf{0.5434} \\ \small{(+78.05\%)}\end{tabular} & 
\begin{tabular}[c]{@{}c@{}}\textbf{0.6835} \\ \small{(+46.86\%)}\end{tabular} & 
\begin{tabular}[c]{@{}c@{}}\textbf{0.7584} \\ \small{(+32.38\%)}\end{tabular} & 
\begin{tabular}[c]{@{} c@ {}}\textbf{0.4966} \\ \small{(+84.54\%)}\end{tabular} & 
\begin{tabular}[c]{@{} c@ {}}\textbf{0.5170} \\ \small{(+75.25\%)}\end{tabular} & 
\begin{tabular}[c]{@{} c@ {}}\textbf{0.6476} \\ \small{(+42.11\%)}\end{tabular} & 
\begin{tabular}[c]{@{} c@ {}}\textbf{0.7280} \\ \small{(+30.56\%)}\end{tabular} \\
\hline
\multirow{7}{*}{\begin{tabular}[c]{@{}l@{}}\textbf{Negative}\\ (\textbf{N = 10})\end{tabular}} 
& {Base} & 0.1287 & 0.1758 & 0.1217 & 0.2070 & 0.1069 & 0.1487 & 0.1019 & 0.1776 \\
& \textit{+ f} & 0.1385 & 0.1865 & 0.1292 & 0.2159 & 0.1130 & 0.1561 & 0.1077 & 0.1585 \\
& \textit{+ f + a} & 0.1378 & 0.1822 & 0.1275 & 0.2075 & 0.1274 & 0.1715 & 0.1182 & 0.1981 \\
\cline{2-10}
& {Filter} & 0.1571 & 0.2097 & 0.1460 & 0.2411 & 0.1309 & 0.1776 & 0.1229 & 0.2075 \\
& \textit{+ f} & \textbf{0.1684} & \textbf{0.2211} & \textbf{0.1551} & \textbf{0.2502} & 0.1382 & 0.1878 & 0.1295 & 0.2195 \\
& \textit{+ f + a} & 0.1601 & 0.2088 & 0.1459 & 0.2341 & 0.1463 & 0.1934 & 0.1349 & 0.2199 \\
\cline{2-10}
& {DPR} & 
\begin{tabular}[c]{@{}c@{}}0.1670 \\ \small{(-0.83\%)}\end{tabular} & 
\begin{tabular}[c]{@{}c@{}}0.2191 \\ \small{(-0.90\%)}\end{tabular} & 
\begin{tabular}[c]{@{}c@{}}0.1537 \\ \small{(-0.90\%)}\end{tabular} & 
\begin{tabular}[c]{@{}c@{}}0.2479 \\ \small{(-0.92\%)}\end{tabular} & 
\begin{tabular}[c]{@{}c@{}}\textbf{0.1609} \\ \small{(+9.98\%)}\end{tabular} & 
\begin{tabular}[c]{@{}c@{}}\textbf{0.2132} \\ \small{(+10.24\%)}\end{tabular} & 
\begin{tabular}[c]{@{}c@{}}\textbf{0.1476} \\ \small{(+9.41\%)}\end{tabular} & 
\begin{tabular}[c]{@{}c@{}}\textbf{0.2420} \\ \small{(+10.05\%)}\end{tabular} \\
\hline
\end{tabular}
\begin{tablenotes}
    \footnotesize \item[1] Relative improvements for DPR are calculated based on the strongest \texttt{Filter} baseline in each sub-task. Bold indicates the best overall performance.
\end{tablenotes}
\end{table*}
\begin{table*}[htbp!]
\centering
\small
\caption{Performance comparison on the MIND dataset with different $N$ settings.}
\vspace{-10pt}
\label{tab:mind_topk}
\begin{tabular}{@{}llcccc|cccc@{}}
\hline
\multirow{2}{*}{\textbf{Task Type}} & \multirow{2}{*}{\textbf{Model}} & \multicolumn{4}{c|}{\textbf{SASRec}} & \multicolumn{4}{c}{\textbf{GRU4Rec}} \\
\cline{3-10}
& & {NDCG@10} & {NDCG@20} & {Recall@10} & {Recall@20} & {NDCG@10} & {NDCG@20} & {Recall@10} & {Recall@20} \\ 
\hline
\multirow{7}{*}{\begin{tabular}[c]{@{}l@{}}\textbf{Positive}\\ (\textbf{N = 3})\end{tabular}} 
& {Base} & 0.0431 & 0.0545 & 0.0835 & 0.1291 & 0.0395 & 0.0504 & 0.0767 & 0.1200 \\
& \textit{+ f} & 0.0441 & 0.0564 & 0.0852 & 0.1341 & 0.0390 & 0.0499 & 0.0759 & 0.1192 \\
& \textit{+ f + a} & 0.0622 & 0.0777 & 0.1200 & 0.1819 & 0.0525 & 0.0665 & 0.1017 & 0.1574 \\
\cline{2-10}
& {Filter} & 0.3259 & 0.3413 & 0.4893 & 0.5496 & 0.3170 & 0.3320 & 0.4786 & 0.5370 \\
& \textit{+ f} & 0.3386 & 0.3542 & 0.5075 & 0.5686 & 0.3117 & 0.3266 & 0.4696 & 0.5270 \\
& \textit{+ f + a} & 0.4127 & 0.4291 & \textbf{0.6030} & \textbf{0.6670} & 0.3798 & 0.3965 & 0.5613 & 0.6265 \\
\cline{2-10}
& {DPR} & 
\begin{tabular}[c]{@{}c@{}}\textbf{0.4679} \\ \small{(+13.38\%)}\end{tabular} & 
\begin{tabular}[c]{@{}c@{}}\textbf{0.4836} \\ \small{(+12.70\%)}\end{tabular} & 
\begin{tabular}[c]{@{}c@{}}0.5998 \\ \small{(-0.53\%)}\end{tabular} & 
\begin{tabular}[c]{@{}c@{}}0.6619 \\ \small{(-0.76\%)}\end{tabular} & 
\begin{tabular}[c]{@{} c@ {}}\textbf{0.4956} \\ \small{(+30.49\%)}\end{tabular} & 
\begin{tabular}[c]{@{} c@ {}}\textbf{0.5117} \\ \small{(+29.05\%)}\end{tabular} & 
\begin{tabular}[c]{@{} c@ {}}\textbf{0.6326} \\ \small{(+12.70\%)}\end{tabular} & 
\begin{tabular}[c]{@{} c@ {}}\textbf{0.6959} \\ \small{(+11.08\%)}\end{tabular} \\
\hline
\multirow{7}{*}{\begin{tabular}[c]{@{}l@{}}\textbf{Negative}\\ (\textbf{N = 3})\end{tabular}} 
& {Base} & 0.0600 & 0.0764 & 0.0827 & 0.1291 & 0.0557 & 0.0711 & 0.0767 & 0.1202 \\
& \textit{+ f} & 0.0620 & 0.0793 & 0.0854 & 0.1343 & 0.0548 & 0.0701 & 0.0758 & 0.1190 \\
& \textit{+ f + a} & 0.0877 & 0.1100 & 0.1202 & 0.1830 & 0.0737 & 0.0935 & 0.1013 & 0.1572 \\
\cline{2-10}
& {Filter} & 0.0643 & 0.0809 & 0.0875 & 0.1345 & 0.0596 & 0.0753 & 0.0809 & 0.1253 \\
& \textit{+ f} & 0.0665 & 0.0841 & 0.0904 & 0.1400 & 0.0588 & 0.0744 & 0.0802 & 0.1242 \\
& \textit{+ f + a} & \textbf{0.0932} & \textbf{0.1159} & \textbf{0.1260} & \textbf{0.1903} & 0.0783 & 0.0984 & 0.1063 & 0.1632 \\
\cline{2-10}
& {DPR} & 
\begin{tabular}[c]{@{}c@{}}0.0868 \\ \small{(-6.87\%)}\end{tabular} & 
\begin{tabular}[c]{@{}c@{}}0.1079 \\ \small{(-6.90\%)}\end{tabular} & 
\begin{tabular}[c]{@{}c@{}}0.1171 \\ \small{(-7.06\%)}\end{tabular} & 
\begin{tabular}[c]{@{}c@{}}0.1767 \\ \small{(-7.15\%)}\end{tabular} & 
\begin{tabular}[c]{@{}c@{}}\textbf{0.0842} \\ \small{(+7.54\%)}\end{tabular} & 
\begin{tabular}[c]{@{}c@{}}\textbf{0.1048} \\ \small{(+6.50\%)}\end{tabular} & 
\begin{tabular}[c]{@{}c@{}}\textbf{0.1136} \\ \small{(+6.87\%)}\end{tabular} & 
\begin{tabular}[c]{@{}c@{}}\textbf{0.1717} \\ \small{(+5.21\%)}\end{tabular} \\
\hline
\hline
\multirow{7}{*}{\begin{tabular}[c]{@{}l@{}}\textbf{Positive}\\ (\textbf{N = 5})\end{tabular}} 
& {Base} & 0.0380 & 0.0493 & 0.0743 & 0.1192 & 0.0361 & 0.0464 & 0.0706 & 0.1119 \\
& \textit{+ f} & 0.0396 & 0.0514 & 0.0782 & 0.1252 & 0.0355 & 0.0456 & 0.0699 & 0.1100 \\
& \textit{+ f + a} & 0.0560 & 0.0711 & 0.1093 & 0.1694 & 0.0470 & 0.0601 & 0.0915 & 0.1438 \\
\cline{2-10}
& {Filter} & 0.3139 & 0.3298 & 0.4775 & 0.5396 & 0.3074 & 0.3228 & 0.4692 & 0.5294 \\
& \textit{+ f} & 0.3279 & 0.3440 & 0.4974 & 0.5605 & 0.2991 & 0.3141 & 0.4545 & 0.5134 \\
& \textit{+ f + a} & 0.3994 & 0.4159 & \textbf{0.5898} & \textbf{0.6542} & 0.3636 & 0.3806 & 0.5431 & 0.6093 \\
\cline{2-10}
& {DPR} & 
\begin{tabular}[c]{@{}c@{}}\textbf{0.4503} \\ \small{(+12.74\%)}\end{tabular} & 
\begin{tabular}[c]{@{}c@{}}\textbf{0.4665} \\ \small{(+12.17\%)}\end{tabular} & 
\begin{tabular}[c]{@{}c@{}}0.5825 \\ \small{(-1.24\%)}\end{tabular} & 
\begin{tabular}[c]{@{}c@{}}0.6468 \\ \small{(-1.13\%)}\end{tabular} & 
\begin{tabular}[c]{@{} c@ {}}\textbf{0.4764} \\ \small{(+31.02\%)}\end{tabular} & 
\begin{tabular}[c]{@{} c@ {}}\textbf{0.4930} \\ \small{(+29.53\%)}\end{tabular} & 
\begin{tabular}[c]{@{} c@ {}}\textbf{0.6149} \\ \small{(+13.22\%)}\end{tabular} & 
\begin{tabular}[c]{@{} c@ {}}\textbf{0.6804} \\ \small{(+11.67\%)}\end{tabular} \\
\hline
\multirow{7}{*}{\begin{tabular}[c]{@{}l@{}}\textbf{Negative}\\ (\textbf{N = 5})\end{tabular}} 
& {Base} & 0.0639 & 0.0824 & 0.0738 & 0.1173 & 0.0608 & 0.0784 & 0.0702 & 0.1114 \\
& \textit{+ f} & 0.0667 & 0.0862 & 0.0772 & 0.1229 & 0.0595 & 0.0768 & 0.0689 & 0.1094 \\
& \textit{+ f + a} & 0.0944 & 0.1200 & 0.1084 & 0.1685 & 0.0789 & 0.1010 & 0.0911 & 0.1430 \\
\cline{2-10}
& {Filter} & 0.0686 & 0.0873 & 0.0782 & 0.1223 & 0.0651 & 0.0829 & 0.0741 & 0.1161 \\
& \textit{+ f} & 0.0716 & 0.0914 & 0.0818 & 0.1282 & 0.0638 & 0.0813 & 0.0730 & 0.1142 \\
& \textit{+ f + a} & \textbf{0.1002} & \textbf{0.1264} & \textbf{0.1137} & \textbf{0.1752} & 0.0837 & 0.1063 & 0.0956 & 0.1486 \\
\cline{2-10}
& {DPR} & 
\begin{tabular}[c]{@{}c@{}}0.0931 \\ \small{(-7.09\%)}\end{tabular} & 
\begin{tabular}[c]{@{}c@{}}0.1170 \\ \small{(-7.44\%)}\end{tabular} & 
\begin{tabular}[c]{@{}c@{}}0.1055 \\ \small{(-7.21\%)}\end{tabular} & 
\begin{tabular}[c]{@{}c@{}}0.1616 \\ \small{(-7.76\%)}\end{tabular} & 
\begin{tabular}[c]{@{}c@{}}\textbf{0.0894} \\ \small{(+6.81\%)}\end{tabular} & 
\begin{tabular}[c]{@{}c@{}}\textbf{0.1128} \\ \small{(+6.11\%)}\end{tabular} & 
\begin{tabular}[c]{@{}c@{}}\textbf{0.1012} \\ \small{(+5.86\%)}\end{tabular} & 
\begin{tabular}[c]{@{}c@{}}\textbf{0.1559} \\ \small{(+4.91\%)}\end{tabular} \\
\hline
\end{tabular}
\begin{tablenotes}
    \footnotesize \item[1] \textit{+ f} and \textit{+ f + a} represent different fine-tuning stages. DPR's improvements are calculated relative to the \texttt{Filter + f + a} baseline. Bold numbers indicate the best results within each sub-task.
\end{tablenotes}
\end{table*}

\subsection{Results for Ablation on Stage Design}
Here we report the results discussed in Section~\ref{sec:ablation_stage}. The result is shown in Table~\ref{tab:stage_ablation}.

\begin{table*}[!htbp]
\caption{Ablation study of the multi-stage framework (3-Stage vs. 2-Stage) on ML-1M and MIND datasets. }
\vspace{-10pt}
\label{tab:stage_ablation}
\centering
\resizebox{\textwidth}{!}{%
\begin{tabular}{cc|ccc|ccc|ccc|ccc}
\hline
    \multicolumn{2}{c|}{Dataset} & \multicolumn{6}{c|}{\textbf{ML-1M}} & \multicolumn{6}{c}{\textbf{MIND}} \\
\hline
    \multicolumn{2}{c|}{Model} & \multicolumn{3}{c|}{\textbf{SASRec}} & \multicolumn{3}{c|}{\textbf{GRU4Rec}} & \multicolumn{3}{c|}{\textbf{SASRec}} & \multicolumn{3}{c}{\textbf{GRU4Rec}} \\
\hline
    \multicolumn{2}{c|}{Stage} & 3-S & 2-S & Diff & 3-S & 2-S & Diff & 3-S & 2-S & Diff & 3-S & 2-S & Diff \\
\hline
    \multirow{4}{*}{\textbf{[SEQ]}} 
    & N@10 & \textbf{0.1838} & 0.1822 & -0.87\% & \textbf{0.1817} & 0.1718 & -5.45\% & \textbf{0.1117} & 0.1093 & -2.15\% & 0.0959 & 0.0959 & 0.00\% \\
    & N@20 & \textbf{0.2161} & 0.2135 & -1.20\% & \textbf{0.2145} & 0.2024 & -5.64\% & \textbf{0.1333} & 0.1310 & -1.73\% & 0.1149 & \textbf{0.1150} & +0.09\% \\
    & R@10 & \textbf{0.3341} & 0.3306 & -1.05\% & \textbf{0.3220} & 0.3079 & -4.38\% & \textbf{0.2022} & 0.1986 & -1.78\% & 0.1745 & \textbf{0.1750} & +0.29\% \\
    & R@20 & \textbf{0.4621} & 0.4541 & -1.73\% & \textbf{0.4520} & 0.4290 & -5.09\% & \textbf{0.2878} & 0.2849 & -1.01\% & 0.2498 & \textbf{0.2511} & +0.52\% \\
\hline
    \multirow{4}{*}{\textbf{[POS]}} 
    & N@10 & \textbf{0.5748} & 0.5297 & -7.85\% & \textbf{0.5501} & 0.4925 & -10.47\% & \textbf{0.4591} & 0.4181 & -8.93\% & \textbf{0.4688} & 0.4597 & -1.94\% \\
    & N@20 & \textbf{0.5921} & 0.5483 & -7.40\% & \textbf{0.5683} & 0.5116 & -9.98\% & \textbf{0.4751} & 0.4334 & -8.78\% & \textbf{0.4855} & 0.4746 & -2.25\% \\
    & R@10 & \textbf{0.7300} & 0.6809 & -6.73\% & \textbf{0.6991} & 0.6414 & -8.25\% & \textbf{0.5912} & 0.5389 & -8.85\% & \textbf{0.6058} & 0.5792 & -4.39\% \\
    & R@20 & \textbf{0.7981} & 0.7543 & -5.49\% & \textbf{0.7710} & 0.7169 & -7.02\% & \textbf{0.6543} & 0.5997 & -8.34\% & \textbf{0.6719} & 0.6383 & -5.00\% \\
\hline
    \multirow{4}{*}{\textbf{[NEG]}} 
    & N@10 & \textbf{0.1772} & 0.1537 & -13.26\% & \textbf{0.1741} & 0.1511 & -13.21\% & \textbf{0.0899} & 0.0884 & -1.67\% & \textbf{0.0876} & 0.0824 & -5.94\% \\
    & N@20 & \textbf{0.2250} & 0.2001 & -11.07\% & \textbf{0.2220} & 0.1963 & -11.58\% & \textbf{0.1124} & 0.1111 & -1.16\% & \textbf{0.1109} & 0.1049 & -5.41\% \\
    & R@10 & \textbf{0.2076} & 0.1832 & -11.75\% & \textbf{0.2030} & 0.1796 & -11.53\% & \textbf{0.1113} & 0.1098 & -1.35\% & \textbf{0.0981} & 0.0929 & -5.30\% \\
    & R@20 & \textbf{0.3172} & 0.2899 & -8.61\% & \textbf{0.3124} & 0.2835 & -9.25\% & \textbf{0.1691} & 0.1682 & -0.53\% & \textbf{0.1513} & 0.1444 & -4.56\% \\
\hline
\end{tabular}%
}
\end{table*}

\subsection{Results for Ablation on Loss Design}
Here we report the results discussed in Section~\ref{sec:ablation_loss}. The result is shown in Table~\ref{tab:loss_ablation}.

\begin{table*}[!htbp]
\caption{Ablation study of Loss Functions on ML-1M and MIND datasets.}
\vspace{-10pt}
\label{tab:loss_ablation}
\centering
\resizebox{\textwidth}{!}{%
\begin{tabular}{cc|ccc|ccc|ccc|ccc}
\hline
    \multicolumn{2}{c|}{Dataset} & \multicolumn{6}{c|}{\textbf{ML-1M}} & \multicolumn{6}{c}{\textbf{MIND}} \\
\hline
    \multicolumn{2}{c|}{Model} & \multicolumn{3}{c|}{\textbf{SASRec}} & \multicolumn{3}{c|}{\textbf{GRU4Rec}} & \multicolumn{3}{c|}{\textbf{SASRec}} & \multicolumn{3}{c}{\textbf{GRU4Rec}} \\
\hline
    \multicolumn{2}{c|}{Loss Type} & Full & Pos & Neg & Full & Pos & Neg & Full & Pos & Neg & Full & Pos & Neg \\
\hline
    \multirow{4}{*}{\textbf{[POS]}} 
    & N@10 & \textbf{0.5748} & 0.5672 & 0.1670 & \textbf{0.5501} & 0.5496 & 0.1630 & 0.4591 & \textbf{0.4629} & 0.0946 & 0.4688 & \textbf{0.5068} & 0.1211 \\
    & N@20 & \textbf{0.5921} & 0.5853 & 0.1973 & \textbf{0.5683} & 0.5672 & 0.1913 & 0.4751 & \textbf{0.4789} & 0.1110 & 0.4855 & \textbf{0.5206} & 0.1373 \\
    & R@10 & \textbf{0.7300} & 0.7219 & 0.3079 & \textbf{0.6991} & 0.6986 & 0.2957 & 0.5912 & \textbf{0.5955} & 0.1633 & 0.6058 & \textbf{0.6292} & 0.1997 \\
    & R@20 & \textbf{0.7981} & 0.7930 & 0.4281 & \textbf{0.7710} & 0.7680 & 0.4079 & 0.6543 & \textbf{0.6588} & 0.2286 & 0.6719 & \textbf{0.6837} & 0.2643 \\
\hline
    \multirow{4}{*}{\textbf{[NEG]}} 
    & N@10 & \textbf{0.1772} & 0.1634 & \textbf{0.1772} & \textbf{0.1741} & 0.1639 & 0.1733 & \textbf{0.0899} & 0.0759 & 0.0896 & \textbf{0.0876} & 0.0831 & 0.0850 \\
    & N@20 & \textbf{0.2250} & 0.2089 & 0.2246 & \textbf{0.2220} & 0.2088 & 0.2216 & \textbf{0.1124} & 0.0956 & 0.1122 & \textbf{0.1109} & 0.1057 & 0.1077 \\
    & R@10 & \textbf{0.2076} & 0.1904 & 0.2070 & \textbf{0.2030} & 0.1910 & 0.2026 & \textbf{0.1113} & 0.0940 & 0.1097 & \textbf{0.0981} & 0.0934 & 0.0954 \\
    & R@20 & \textbf{0.3172} & 0.2947 & 0.3156 & \textbf{0.3124} & 0.2938 & 0.3131 & \textbf{0.1691} & 0.1446 & 0.1679 & \textbf{0.1513} & 0.1450 & 0.1474 \\
\hline
\end{tabular}%
}
\end{table*}

\subsection{Results for Ablation on Tower Architecture Design}
Here we report the results discussed in Section~\ref{sec:ablation_tower}. The result is shown in Table~\ref{tab:tower_ablation}.

\begin{table*}[!htbp]
\caption{Ablation study of Model Architecture (Two-Tower vs. Single-Tower) on ML-1M and MIND datasets.}
\vspace{-10pt}
\label{tab:tower_ablation}
\centering
\resizebox{\textwidth}{!}{%
\begin{tabular}{cc|ccc|ccc|ccc|ccc}
\hline
    \multicolumn{2}{c|}{Dataset} & \multicolumn{6}{c|}{\textbf{ML-1M}} & \multicolumn{6}{c}{\textbf{MIND}} \\
\hline
    \multicolumn{2}{c|}{Model} & \multicolumn{3}{c|}{\textbf{SASRec}} & \multicolumn{3}{c|}{\textbf{GRU4Rec}} & \multicolumn{3}{c|}{\textbf{SASRec}} & \multicolumn{3}{c}{\textbf{GRU4Rec}} \\
\hline
    \multicolumn{2}{c|}{Arch.} & 2-T & 1-T & Diff & 2-T & 1-T & Diff & 2-T & 1-T & Diff & 2-T & 1-T & Diff \\
\hline
    \multirow{4}{*}{\textbf{[POS]}} 
    & N@10 & \textbf{0.5748} & 0.4173 & -27.40\% & \textbf{0.5501} & 0.3941 & -28.36\% & \textbf{0.4591} & 0.3158 & -31.21\% & \textbf{0.4688} & 0.3666 & -21.80\% \\
    & N@20 & \textbf{0.5921} & 0.4396 & -25.76\% & \textbf{0.5683} & 0.4164 & -26.73\% & \textbf{0.4751} & 0.3329 & -29.93\% & \textbf{0.4855} & 0.3816 & -21.40\% \\
    & R@10 & \textbf{0.7300} & 0.5650 & -22.60\% & \textbf{0.6991} & 0.5390 & -22.90\% & \textbf{0.5912} & 0.4288 & -27.47\% & \textbf{0.6058} & 0.4755 & -21.51\% \\
    & R@20 & \textbf{0.7981} & 0.6528 & -18.21\% & \textbf{0.7710} & 0.6270 & -18.68\% & \textbf{0.6543} & 0.4963 & -24.15\% & \textbf{0.6719} & 0.5348 & -20.40\% \\
\hline
    \multirow{4}{*}{\textbf{[NEG]}} 
    & N@10 & \textbf{0.1772} & 0.1146 & -35.33\% & \textbf{0.1741} & 0.1152 & -33.83\% & \textbf{0.0899} & 0.0507 & -43.60\% & \textbf{0.0876} & 0.0477 & -45.55\% \\
    & N@20 & \textbf{0.2250} & 0.1581 & -29.73\% & \textbf{0.2220} & 0.1592 & -28.29\% & \textbf{0.1124} & 0.0707 & -37.10\% & \textbf{0.1109} & 0.0688 & -37.96\% \\
    & R@10 & \textbf{0.2076} & 0.1487 & -28.37\% & \textbf{0.2030} & 0.1484 & -26.90\% & \textbf{0.1113} & 0.0727 & -34.68\% & \textbf{0.0981} & 0.0642 & -34.56\% \\
    & R@20 & \textbf{0.3172} & 0.2490 & -21.50\% & \textbf{0.3124} & 0.2501 & -19.94\% & \textbf{0.1691} & 0.1241 & -26.61\% & \textbf{0.1513} & 0.1128 & -25.45\% \\
\hline
\end{tabular}%
}
\end{table*}

\end{document}